# A perspective on the Fe-based superconductors

John A. Wilson

H.H. Wills Physics Laboratory
University of Bristol
Bristol BS8 1TL  U.K.

**Abstract**

FeSe is employed as reference material to elucidate the observed high $T_c$ superconducting behaviour of the related layered iron pnictides.  The structural and ensuing semimetallic band structural forms are here rather unusual, with the resulting ground state details extremely sensitive to the precise shape of the Fe-X coordination unit.  The superconductivity is presented as coming from a combination of Resonant Valence Bond (RVB) and Excitonic Insulator physics, and incorporating Boson-Fermion degeneracy.  Although sourced in a very different fashion the latter leads to some similarites with the high temperature superconducting (HTSC) cuprates.  The Excitonic Insulator behaviour sees spin density wave, charge density wave/periodic structural distortion (SDW, CDW/PLD), and superconductive instabilities all vie for ground state status.  The conflict leads to a very sensitive and complex set of properties, frequently mirroring HTSC cuprate behaviour.  The delicate balance between ground states is made particularly difficult to unravel by the micro-inhomogeneity of structural form which it can engender.  It is pointed out that several other notable superconductors, layered in form, semimetallic with indirect overlap and possessing homopolar bonding, would look to fall into the same general category, $\beta$-ZrNCl and MgB$_2$ and the high-pressure forms of several elements, like sulphur, phosphorus, lithium and calcium, being cases in point.

## §1. Contents and Introduction.

It had been my intent not to become involved with the pnictide superconductors, which were taken as a distraction from the hole-doped cuprate superconductors, and which it would appear are in another class, another league. However the pnictides are beginning to project certain lines of thought, detrimentally I believe, back into the cuprate problem. Moreover preconceptions concerning what is afoot in the pnictides seem to be taking many in altogether the wrong direction in that field too. Here I draw upon my long background [1,2,3,4] in addressing the full array of transition metal materials, often from a somewhat more chemically oriented perspective than the majority of solid state physicists, in order to (*1*) present the critical situation in play in the pnictide systems, (*2*) relate it to the different situation existing in the cuprates, and (*3*) indicate how it will generalize to a striking class of newly emergent superconductors, namely the semimetallic, homopolar-bonded, low-dimensional polymorphs of the elements, among them iodine, phosphorus, sulphur, lithium, calcium, yttrium, etc., accessed under high pressures.

I have long discussed the mixed-valent HTSC cuprates in terms of negative-$U$ effects (based on $p^6d^{10}$ shell closure within a Cu(III) coordination unit), the superconductivity being then driven by boson-fermion resonant crossover. The touchstone to such behaviour lies with bulk disproportionation. Conversely, with the pnictides the corresponding bases are perceived as being Resonant Valence Bond (RVB) behaviour and the Excitonic Insulator. As with the HTSC cuprates, the striking superconductive properties are once again viewed as arising in a resonant mixture of bosons and fermions, but this time gained under novel semi-metallic governance.

In what follows

§2 surveys the chemical and structural setting of mackinowite FeS and FeSe, emphasizing the control the unusual bisphenoidal coordination geometry exercises over the clash between the magnetic and superconductive behaviour exhibited here and in the related pnictides.

§3 provides a closer look at the very profound effect that direct Fe-Fe bonding has upon the sequencing of the various d-states within the present $Z$=2 structural setting.

§4 makes an examination of the resulting semimetallic band structure and the critical role that correlation here has in arriving at an adequate description of the experimental findings.

§5 looks at the intertwining of Resonant Valence Bond and Excitonic Insulator physics in establishing high $T_c$ superconductivity here. Exposition is made of the similarities and differences between what is occurring in the pnictides as against in the mixed-valent HTSC cuprates.

§6 takes a detailed look at ten very different types of experimental work reported for the pnictides and interprets the results on the above basis.

§7 examines how widely-based adjustment to the chemical content of the unit cell facilitates a very sensitive and complex control over the interplay between the magnetism and the superconductivity observed with this group of materials.

§8 points to the important role that the lattice exercises within the tripartite action precipitated by the Excitonic Insulator/RVB physics in play.

§9 probes the form and origin of the electronic and structural inhomogeneities generated in the delicate balance between possible ground states for the present semimetallic systems.

§10 suggests that the superconductivity at elevated temperatures met with in certain other low-dimensional, semimetallic materials that involve homopolar bonding, such as many non-metallic and metallic elements under high pressure, appears to partake of certain aspects of what is arising in the pnictides, as likewisedo materials based on β-ZrNCl.

§11 attempts to draw together all the many strands included in this survey and to restate the form of interpretation being offered for pnictide superconductivity, making contrast with what has been set out previously concerning cuprate HTSC behaviour.

**§2. A background to the pnictide materials.**

FeAs is a magnetic, metallic material, but it is an Fe(III) material, and the pnictide superconductors we are dealing with here are Fe(II). LnO.FeAs basically is a tetragonal layered form of FeAs fully intercalated by the monovalent cation, LnO$^+$. (N.B. the Ln atom sits on the outside of the LnO sandwich). The uranyl (UO$_2$)$^{2+}$ and vanadyl (VO)$^{3+}$ cations are further, perhaps more familiar, examples of such oxycations, common for the more electropositive elements. The present (LnO)$^{1+}$ role [5] becomes explicit with the comparable system LiFeAs ($T_c$ = 21 K) [6], and indeed BaFe$_2$As$_2$ [7] once written Ba$_{½}$.FeAs. (Note the latter under pressure becomes superconducting without any recourse to potassium doping [8]). Now one's frame of reference in contemplating the properties of NaTiO$_2$ is not Ti(IV) but Ti(III), and in the present case it is not Fe(III) but Fe(II). Accordingly it is to FeSe, not FeAs, that one is directed. It has by now become well publicized that tetragonal FeSe exhibits the same characteristic superconductive aspect to its behaviour, particularly under pressure [9], as that found in the more widely researched pnictides. Other crystallographic forms of FeSe do not display anything comparable, and hence the key to the behaviour has to lie with the detailed crystal structure and ensuing band structure.

The highly unusual nature of the tetragonal Mackinowite form of FeS has long been recognized [10,2]. FeS normally is encountered in hexagonal nickel arsenide form (Pyrrhotite, toward which Mackinowite is unstable above 150°C). In Pyrrhotite the iron ions are octahedrally coodinated and show high spin (h.s.) behaviour to produce an $S$ = 2 magnetic metallic material [11]. At lower temperature in stoichiometric samples of pyrrhotite a basal trimerization occurs, seen in the mineral form Troilite [12]. FeS finally can be secured in tetrahedrally coordinated zincblende form [13]. With the latter there exist no short Fe-Fe distances and this polymorph manages to retain Mott insulating character, much as in octahedrally coordinated h.s. $d^6$ FeO. Remember, though, now in tetrahedral coordination – at least given a regular tetrahedron in apical orientation – the ten $d$-states will experience local symmetry splitting 4-and-6 ($e/t_2$), not the 6-and-4 $t_{2g}/e_g$ division so familiar with octahedral coordination [14].

In the mackinowite form of FeS the structure displays several important differences from its Mott-insulating zincblende polymorph. The tetrahedra now share edges rather than corners [10]. That brings about a layered tetragonal 2D net rather than a 3D one, and one most importantly containing a much shorter nearest-neighbour Fe-Fe distance (2.60 Å vs 3.83 Å). The tetrahedra no longer are constrained by the adopted structure to be ideal in form (i.e. of uniform interbond angle 109° 28'), and they find themselves not in apical orientation either. The tetrahedra in this uncommon tetragonal structure form two mutually perpendicular, basal strings of bisphenoids (crossed wedges) with the crystallographic $z$ axis running between opposed edge-centring locations. All the iron atoms reside at crystallographically equivalent sites (point symmetry $\bar{4}$2m), the unit cell holding two such atoms (i.e. $Z$ = 2). The space group is No. 129, non-symmorphic $D_{4h}^7$ (P4/nmm). This implies for the band structure the presence of twice as many bands as with the $Z$ = 1 zincblende polymorph (e.g. 6 sulphur-based *p*-bands, not 3) – a feature shortly to prove key in understanding how mackinowite FeS ends up a semiconductor.

Now mackinowite FeS is not the only semiconducting sulphide of Fe(II). The best known of all iron sulphides is iron pyrites (see figure 7 in [1]). $FeS_2$ derives from rocksalt FeO when all the $O^{2-}$ ions are replaced by $(S_2)^{2-}$ pseudo-halogen dimers. These dimer ions align systematically along the four different body diagonals of the cubic structure reducing the space group from $O_h^5$ (Fmmm) to $T_h^6$ (Pa3). In the rocksalt structure the octahedra shared all their edges, but now in the pyrite structure the $S_2^{2-}$ units spring the octahedra apart so as to share corners only. The outcome is a $Z$ = 1 basis with f.c.c. rocksalt, but with the pyrite-structured material a $Z$ = 4 basis. Transfer from the oxide to the sulphide moreover tips the balance from a high-spin to a low-spin state; i.e. the increased covalence (*p/d* hybridization) of the sulphide causes crystal-field/molecular-orbital effects to become dominant over Hund's rule spin coupling. One can in fact follow such transfer under pressure in hexagonal FeS [15]. Crossover occurs quite rapidly as a result of the sharp reduction in bond length (and hence cell volume) accompanying conversion of antibonding $dp\sigma^*$ (here $e_g$) electrons into non-bonding $dp\pi$ (here $t_{2g}$) electrons. With $d^6$ $FeS_2$ the end product is a minimum semiconductive band gap of 0.9 eV between the $t_{2g}$ and $e_g$ manifolds [16]. Since in $FeS_2$ $Z$ = 4, there are going to exist 3x4 = 12 of the former subbands and 2x4 = 8 of the latter in the developing horizontal spaghetti.

With $FeS_2$ and the rest of the 3*d* pyrite disulphides, ($d^7$) $e_g^1$ $CoS_2$, ($d^8$) $e_g^2$ $NiS_2$, ($d^9$) $e_g^3$ $CuS_2$, and ($d^{10}$) $e_g^4$ $ZnS_2$, great simplicity is conferred upon their properties by the cations remaining tethered to the f.c.c. sublattice. This is what makes $NiS_2$ the archetypal material for studying the Mott transition [17,1,4], as against a corundum-structured material like $d^2$ $V_2O_3$, where the cations are free to shift along the *c*-axis – indeed in $d^1$ $Ti_2O_3$ to dimerize. In the alternative semiconducting, orthorhombic, Marcasite polymorph of $FeS_2$ (see figures 8 and 9 in [1]) the octahedra there now form edge-sharing strings. The capacity then for cation dimerization becomes realized under the appropriate $t_{2g}^5$ circumstance of arsenopyrite Fe(AsSe), this negated upon returning to $t_{2g}^6$ and Co(AsSe). When dealing with all these materials it is essential always to appreciate the functionality of the component units; witness $Fe_2(P_2S_6)$ is a water-soluble thiophosphite [18], not a phospho-sulphide.

One is required now to address how mackinowite FeS manages to be a semiconductor at $d^6$ when holding *tetrahedral*, not octahedral coordination. This non-metallic outcome is accomplished without any crystallographic distortion from the simple layered tetragonal structure, one shared by Li(OH) – and in anti-site form by PbO. A low-spin non-magnetic condition immediately is made evident from the single-spiked Mössbauer spectrum [10,19]. That however does not simply imply core diamagnetism, as with a standard non-T.M. semiconductor. A strong, countering, almost temperature independent van Vleck paramagnetic term is found here [20], much as in $FeS_2$ [21]. One accepts non-metallic $FeS_2$ to be a semiconductor and not a Mott insulator since it can readily be doped both *n*-type and *p*-type to yield material of moderately high mobility (≥ 1 $cm^2$/V-sec) [1]. Such might not be true of the 3$d^6$ oxide $Co_2O_3$ [22] (*l.s.* analogue of corundum-structured 4$d^6$ $Rh_2O_3$ and perovskite-structured $LaRhO_3$ [23]), but it appears the case for tetragonal FeS. By FeSe there is no doubting its delocalized credentials, by then there occurring significant semi-metallic *d-d* overlap under the augmented covalent *p/d* hybridization. Nonetheless a more customary magnetism emerges in 'FeSe' when one allows the stoichiometry of the system to deviate from strictly 1:1 [24]. Superconductivity in contrast occurs in FeSe under standard conditions right at stoichiometry [25] with $T_c$ = 9 K. From this value $T_c$ proceeds to climb in two stages to a maximum of 36 K [9] under a hydrostatic pressure of 8 GPa (80 kB), before declining again. Clearly the superconductivity showing up here is highly sensitive in kind to the detailed adjustment of the crystallographic free parameter, $u_z$, detailing the precise location of the Se atoms along the *c*-axis (see Figure 2 below); i.e. the superconductivity is highly dependent upon the precise shape of the Fe-Se coordination unit.

In the pure selenide the bisphenoid is slightly elongated along the *c*-axis [24a,26], and, for a layer compound such as the present, *hydrostatic* pressure will actually increase that form of deformation, as the softer van der Waals (vdW) gap region permits the more rigid Fe-Se framework of bonds to concertina into it. Comparable shape sensitivity is apparent with the LnOFeAs family of superconductors, for which the maximum attainable $T_c$ peaks across the lanthanide sequence at Sm, as is seen in figure 1 [27]. With the latter '1111'-materials, *basally directed* chemical stress becomes transmitted from the contracting LnO layers into the FeAs layers as the LnO ionic unit steadily is diminished in size down through the $f^0$ to $f^{14}$ sequence. The stress becomes appropriate within this family optimally to tune $T_c$ to 54 K at the case of $f^5$ Sm(III). And the telling empirical observation here regarding what this tuning amounts to is that it imports to the bisphenoidal FeAs coordination unit ideal tetrahedral geometry: that is the interbond angle As-Fe-As across the *z* axis, $\alpha$, to be referred to as the bond splay angle, takes on the ideal tetrahedral value of 109° 28'. [Note $\alpha = 180°-2\theta$, where $\theta = \tan^{-1}(^{u.c}/_{½a})$ ; see fig. 2 below]. Such ideality of local form is not one enforced by the crystal space group, but one that when attained critically influences the local molecular orbital energies and interactions. The consequences feed through automatically into the detailed band structure to govern the material's electronic properties. Under the structurally resonant conditions, SmOFeAs (as with Mackinowite itself) takes up a non-magnetic, Fe-singlet condition of the same key form, in evidence once again in the Mössbauer and nmr data [28]. Note from figure 1 many of the other materials presently being discussed in connection with pnictide superconductivity, like LaOFeP [29,30], lie with $\alpha$ far removed from ideality, and these systems exhibit varying degrees of resurgent magnetic behaviour and then (much) lower superconducting $T_c$'s.

### §3. From molecular orbitals to a band structure for mackinowite FeS.

Let us first examine in greater detail the form of the Mackinowite structure. Figure 2a shows an ideal tetrahedron in bisphenoidal orientation, with *x*, *y* and *z* axes attached as appropriate to the tetragonal structure. The four triangular faces all are inclined and at mid-height, where the iron lies, they define a horizontal, square intersect, featuring in the basal projection of figure 2b. In tetragonal FeS these tetrahedral coordination units share all four inclined edges to produce a structure rather like a waffle, where square pyramidal pits just extend through the S-Fe-S layering, the pits on the reverse side then being displaced by (½,½) from those on the top. The small Fe ions fill all the bisphenoidal interstices between the S layers to yield a square-centred array and the 1:1 stoichiometry. In FeS the waffles sit directly above one another. Figures 2c and 2d illustrate the situation in side elevation and in plan respectively.

If the van der Waals (vdW) gaps between successive sandwiches were to hold exactly the same interlayer S-S separation as that within the S-Fe-S sandwiches, the S layers would sit at $z$ = ¼ and ¾. However the vdW region is much reduced in thickness. While $u_z(S)$ indeed proves to be somewhat bigger than ¼, this does not immediately imply that the FeS coordination units themselves are automatically stretched along $z$ compared with ideal geometry. Upon using the above quoted formula one discovers that in fact $\alpha$ (the intra-sulphur-sheet bond-splay angle) is remarkably close there to the ideal value of 109.45°. With the selenide, for which the atomic positions are better refined ($u_z$=0.267), $\alpha$ proves to lie appreciably below ideal at 104.1°, meaning the coordination unit now indeed is elongated [24a]. We throughout shall employ angle $\alpha$ to monitor deformation of the tetrahedron, whether as presently or in the related structures of LnOFeAs, BaFe$_2$As$_2$, etc. (for which $\alpha$ most frequently is greater than ideal, as may be seen from figure 1). Given the current layer compound setting, as was indicated above, the initial effect of applied *hydrostatic* pressure [9] is to push up coordination unit elongation, decreasing angle $\alpha$ . The slight (orthorhombic) quadrupolar splitting of the Mössbauer singlet found in FeSe at lower pressure actually vanishes in this process [26c] – a matter of much interest later in §8.

Although the Fe ions within a layer sit on a square-centred array, the FeS sandwich as a whole does not possess face-centred symmetry. Each sandwich of the primitive tetragonal cell contains two Fe, in bisphenoids possessing 90° relative twist (the two otherwise being physically equivalent). This latter twist means the horizontal mirror plane of the relevant space group, P4/nmm, is not a simple mirror but a glide plane associated with translation (½,½). A whole series of vertical glide planes (in 45° orientation) along with horizontal screw diads (both in axial and 45° orientation) result, as set out in the International Crystallographic Tables (Vol.1; group # 129). We have portrayed above in figure 2 what might be termed the natural setting for our material, positioning a Fe atom at the cell corner, and this fortunately proves (in the present case) to be a crystallographically acceptable choice (known as 'setting 1'). However a cell choice displaced from the above by (¼,¼) (what is known as 'setting 2') perhaps better exposes the equivalent disposition of the two Fe atoms within the *Z*=2 structure. The above space group is going automatically to bring about ubiquitous band degeneracy in certain parts of the band structure, specifically within the vertical faces of the Brillouin zone.

Before examining the band structure itself it is beneficial for this material - one still quite close to the Mott transition - to establish a molecular orbital picture of those strong local ingredients set to proceed into the overall crystal potential. One crucial feature to appreciate here is that the natural local symmetry axes of the tetrahedra are not those of the crystal structure $x,y,z$, introduced in figure 2, but $X,Y,Z$, rotated from the former set by 45° in the basal plane. The non-bonding orbitals become labelled then $d_{XY}$ and $d_Z^2$, each pointing towards the edge mid-points of the tetrahedron. $d_{XY}$ accordingly aligns in the directions of the four nearest-neighbour Fe sites within the crystal lattice. These neighbour Fe sites fall so close (at only 2.6 Å in the sulphide) that there will arise significant $dd\pi/\pi^*$ basal interaction in the band deriving from this source (with very little $k_z$ dispersion). By contrast $d_Z^2$ because of its moderate δ-type overlap with the $p_z$ orbitals of the sulphur atoms will display some $k_z$ dispersion, but now with rather little basal dispersion. Of the *anti*bonding $pd\sigma^*$ states, the one parented by $d_{X^2-Y^2}$ engages relatively weakly with all four sulphur atoms and will be least strongly elevated, while the remaining pair of states $d_{XZ}$ and $d_{YZ}$, which rise more strongly in $\sigma^*$ fashion from interaction with the sulphur sublattice, will in addition experience appreciable bonding/antibonding $\pi$ interaction with their n.n. iron atoms (though not quite as much as $d_{XY}$).

In figure 3 we present the evaluated $Z$=2 pattern of M.O. levels; this duly will go forward to dictate Γ-point ordering in the full band structure. The numbers in brackets here mark the state electron complements. The designations applied to the states relate, note, now to the crystallographic axes and *not* to the local axes for the coordination unit introduced above. Behind this transcription will lie appropriately symmetrized state mixing. For FeS one should expect the main valence band, largely parented by the six sulphur 3$p$-states, to lie fully separated off from the $d$-dominated states above. Straightaway from figure 3 one is able to recognize the remarkableness of a semiconducting (or even low-overlap semimetallic) end-product with our current complement of twelve 'd' electrons. It will call for more than simply the weak splitting of the $xy$ level within the $Z$=2 band structure, in view of the simultaneous strong movements of the topmost $x^2-y^2$ ($dd\pi^*$) state upwards and of the lowermost ($dd\pi$) level deriving from the $xz/yz$ doublet downwards. Responsibility for a very low density-of-states outcome at $E_F$ has in some measure to stand as a consequence of the relatively low symmetry of the space group. With representation characters being there restricted over large sections of the zone to unity, such bands are forbidden to cross but necessarily will hybridize and gap. The bands in addition become more readily able to acquire compatible tie-in's in any given energy range. Nonetheless it remains remarkable that the bottommost states from $\sigma^*$ $xy$, as again with the degenerate $\pi$-bonding pair from $yz$ and $zx$, can manage throughout to pick up dispersion to *lower* energy, whilst the elevated $x^2-y^2$ $\pi^*$-state acquires dispersion to *higher* energy, all in a manner such as facilitate the slotting through of a gap embracing precisely the 12 electrons in hand. The role of spin and charge correlation in securing this outcome will need to be addressed further.

Figure 4 provides, without offering much insight into the above matters, a modern LAPW band structure for tetragonal FeS [31]. The shading added here highlights the extensive direct gapping suited to accommodation of our 12 electrons in a low DOS fashion. Whether in reality with the sulphide there is very slight semi-metallic overlap as here, or whether a more refined treatment of the correlation will secure a true semiconducting gap is a problem for later. Where semimetallic overlap definitely is encountered is in the selenide due to the now greater degree of *p*/*d* hybridization, but remarkably the carrier pockets observed experimentally, even with heavy tellurium admixture [32], still appear quite small.

Before closing this section it has to be noted that the above band structure from Subedi and coworkers [31] in fact is not the first to be derived for Mackinowite. On scanning the literature an earlier calculation from Welz and Rosenberg [33] was uncovered, made in the period just prior to HTSC and well before current interest in the pnictide and related superconductors. Their paper ties in nicely much of the early literature on the mackinowite problem. The method they employ is SCLMTO, pursued within the atomic sphere and local density approximations. In its detailed form the product matches remarkably closely the more recent version, thereby serving to emphasize the still rather localized nature of FeS. With tetrahedrally coordinated FeS it is expected from the overview reached spectroscopically and magnetically in [2] that Hubbard *U* will here be someway below the overall *d*-bandwidth, and probably in the range of 1.5 to 2 eV.

One of the benefits of the above LMTO calculation is that it provides valuable information regarding the symmetry representations for the bands at the various special points and along the special lines of the B.Z. That information has been transcribed now from [33] onto the bands of figure 4. The representation labels used follow the notation introduced by Bradley and Cracknell in their extensive group theoretical review [34]. Take note of the universal band degeneracies within the vertical faces of the B.Z. (along XM, MA and AR) instigated by the glide plane in P4/nmm.

## §4. Pnictide band structures and the matter of coordination unit shape tuning for $T_c$ optimization.

The band structure for LaO.FeAs is in general form strikingly similar to that of tetragonal FeSe as can be seen from figure 5. This particular calculation comes from Vildosola *et al* [35] and it was generated using the Full-Potential APW and local orbitals method, and implemented under the WIEN2K code within the LDA correlation and exchange approximation. The band structural detail in close proximity to $E_F$ emerges, as anticipated, as being supersensitive to surprisingly small changes in $u(z)$ and angle $\alpha$. These coordination unit shape changes prove, as evidenced above in figure 1, to be critical for the magnetic/superconducting outcome acquired in any given member of the family. The closer $\alpha$ comes to 109° 28', the more favourable conditions are toward superconductivity and to elevated $T_c$. The situation very graphically is portrayed in figure 6 for the $BaFe_2As_2$ system. Since the latter compound suffers a small orthorhombic distortion at low temperatures which is detrimental to superconductivity, this material customarily is encountered doped with potassium so as to suppress the lattice distortion and grant access to the superconducting state. It is found that there exists an optimum doping level in the latter procedure. With $BaFe_2As_2$ a comparable outcome may, as already noted, also be realized under pressure. A very telling discovery tying the above two eventualities together is that made by Kimber, Kressig, *et al* [36]. Their key observation, reproduced now with some clarification in figure 6, is that, when pursuing *either* procedure, angles $\alpha$ and $\phi$ come to assume tetrahedrally ideal form exactly as $T_c$ optimizes. To understand what is going on here one needs to return to the band structure to see just what such shape ideality crossover actually secures.

Drawing once more on the work of Vildosola *et al* [35], figure 7 presents the detailed form of the band structure in the vicinity of $E_F$ for two contrasting 1111-materials, LaOFeAs ($\alpha$ =113.5°; $T_c^{opt}$ ~26 K) and LaOFeP ($\alpha$ =119°; $T_c^{opt}$ ~6 K). The bands have here been labelled with the dominant state parentage symbols identified in figure 3. Examination of these two differently detailed band structures uncovers three changes of import existing between them. On passing from the phosphide to the arsenide (and to higher $T_c$), one observes (a) a considerably deeper location at $\Gamma$ for $d_z^2$, this lying now well below $E_F$, (b) a considerably lowered energy at $\Gamma$ as well for the (hole-bearing) doublet $d_{xz},d_{yz}$ relative to the conduction band minimum at M, (c) a $d_{xy}$ position not greatly altered with reference to the latter point, this meaning doublet $d_{xz},d_{yz}$ has been brought in the arsenide to a level much closer to degeneracy with the $d_{xy}$ state. The above changes are perceived to issue from the following causes: (a) comes from a reduced broadening of the $d_z^2$ band in the arsenide, which in turn comes from reduction in $d_z^2/p_z$ overlap as the distance to the arsenide layer recedes under the augmented sandwich height with the fall in $\alpha$: (b) is an outcome of the smaller angle $\alpha$ leading, moreover, to greater state separation between the $dd\pi$ and $\pi^*$ states that occur within the $d_{xz}$ and $d_{yz}$-derived set of states, due to the now stronger n.n. Fe interaction: (c) the latter change forces the inter-Fe-site $\pi$-bonded doublet of states ($d_{xz},d_{yz}$) into a closer proximity with the (lower) $d_{xy}$ state – which recall from figure 3 is itself an inter-Fe-site bonded state. The above advance toward full degeneracy of the $d_{xz}$, $d_{yz}$ and $d_{zx}$ states would look to be what is procured as SmOFeAs approaches shape ideality.

Now why should this coincidence have such profound influence upon the superconductivity attained? An important feature to register here is that the three key states figuring above all are spin-paired states, each entailing secondary M-M bonding interaction between near-neighbour Fe sites. They facilitate a prime example of resonant spin-pair (RVB) coupling, sustained for as long as the tetragonal lattice symmetry is retained. An undistorted environment expresses here the widely based covalent hybridization, often further assisted by disorder and the metallicity coming with doping. Breaking this structural symmetry through orthorhombic distortion sees the close degeneracy between the above three orbitals significantly impaired, and this can encourage local moment growth and ordering, always detrimental to superconductivity.

The occurrence of spin coupling in RVB fashion still demands a quite high degree of correlation, and such is precisely the area in which the current calculations are patently somewhat defective. As can be seen from figure 5, the overall $d$-band width that the APW-LDA routine prescribes for LnOFeAs is 4 eV. Given that in this material $U$ is estimated to be ~ 1.5 to 2 eV, it is evident the as-calculated bandwidth is overstating reality by roughly a third. Such renormalization of the calculated bandwidth under more elaborate treatment of local correlation effects than is afforded by the LDA approximation is strongly supported both by ARPES and dHVA/SdH (de Haas-Van Alphen/Shubnikov-de Haas) experimentation. The photoemission results of Li *et al* [37] for LaOFeP would suggest an actual overall *d*-band width of only around 2.5 eV. The quantitative support dHVA provides in this direction [29] comes in the form of deduced band masses enhanced roughly twofold as compared with the corresponding masses gained from the band calculations. To achieve any match between the areal dHVA data and calculation has necessitated slight but significant mutual adjustment of the locations of the occupied and unoccupied bands about $E_F$. Their relative movement lies in the direction of shrinking all the carrier pocket sizes by diminishing semimetallic overlap, the electron band being shifted up, the hole band shifted down each by around 50 meV. In the main these data come, it should be pointed out, from material standing still quite some way from having $\alpha$ right at shape ideality. This is due not only to lack of suitable crystals, say for SmOFeAs, but also because in the dHVA/SdH work if $T_c$ is high, then $H_{c2}$ is high. Accordingly it no longer is possible to sense the normal state given the size of fields available (one requires $H$ to be at least in excess of $H_{irr}$, i.e. running up beyond 30 tesla).

Now with conventional superconductors, and following the MacMillan and Dynes / Eliashberg approach (see Kresin and Wolf [38]), one customarily for high $T_c$ within the BCS scenario looks towards a high density-of-states at $E_F$ in conjunction with a strong electron-phonon coupling parameter, $\lambda$. In the present case neither of these are forthcoming (for analysis of this latter see Boeri *et al* [39]). On the contrary we are afforded prime manifestation of the non-conventional nature of the present superconductors. The current data for shape-optimized pnictides positions them plumb on the Uemura plot [40] for unconventional superconductors – a plot embracing the organic superconductors, the Bucky Ball superconductors, $\beta$-HfNCl, and above all the cuprate HTSC superconductors.

### §5. A perspective of the HTSC mechanism in FeSe and the iron pnictides that avoids over-eager appeal to spin fluctuations.

The picture that we shall pursue below then is essentially the following. In FeSe and its ternary and quaternary analogues (fig. 8) the unusual crystal structure inserts into the band structure a deep, rather narrow, density-of-states minimum at $d^6$. Strong residual correlation effects in the materials - ones not grossly removed from the Mott transition - secure an enhancement of that minimum, to a point where in FeS a semiconducting outcome can just be reached. This correlation effect can be expressed in terms of RVB spin-pair coupling, highly suited to the prevailing situation by virtue of the crystal structure holding two Fe atoms per unit cell, close in separation and in regular, face-centred square, layered setting. As long as the crystal structure holds and does not yield to symmetry distortion, the regular RVB bonding steadily establishes a low susceptibility, paramagnetic environment [41]. The then much constrained spin-flip pair-breaking constitutes a major asset toward the development of superconductivity. With regard however to conventional superconductivity, we pointedly are not in a promising situation since the density of states is very low and the material does not lend itself to strong electron-phonon coupling. Also, as layer compounds, their Debye temperatures are very low too (~ 250 K). What the systems do have running in their favour, however, is that the semimetallic overlap is associated with an indirect gap. (In this they are distinct, say, from the A15 superconductors like $V_3Si$.) With indirect overlap, there emerges for the semimetal a susceptibility to electronic instability with wavevector separating the overlapping electron and hole pockets.

In accord with the excitonic insulator treatment for this form of incipient instability [42], the system is prone either to a spin-density wave, a charge-density wave, or a superconducting instability, or all three, as appropriate to the individual circumstance. Charge- and spin-density waves ordinarily are addressed in terms of extensive and effective Fermi surface 'nesting', examples being Cr and $1T\text{-}TaS_2$ [43]. However with the excitonic insulator situation the Fermi volume is rapidly approaching zero as the gap-bounding bands withdraw their overlap. For the pnictides it actually has become very apparent that the quality of geometrical nesting is not particularly good [29], especially where $\alpha$ remains at a distance from ideality (i.e. from the degeneracy between $d_{xy}$ and $d_{xz}, d_{yz}$), and/or the system has been 'doped', so destroying equality between the number of holes and electrons.

In an excitonic insulator situation one should not look to action of quite the same form as under more extended density-wave interaction. For example the excitonic instability set up in the indirect semimetal $TiSe_2$, and running back someway towards $TiS_2$ [44], brings a state which, although taking on the overlap wavevector $(\pi,\pi,\pi)$, displays an atomic displacement pattern involving not the LA mode, as in a standard CDW, but a TA mode [45,46]. It should be noted here that in density waves of all types the shifts of the atoms from their pre-displaced locations frequently can be appreciably larger than is directly manifest from the lattice parameters themselves. With $2H\text{-}TaS_2$, Ta shifts of 0.1 Å bring only 0.01 Å changes to $a_o$, whilst in $1T\text{-}TaS_2$ with Ta shifts approaching ½ Å even these for a long time escaped being registered. The 'martensitic' changes developed within the $Z$=2 A15 unit cell of $V_3Si$ and $Nb_3Sn$ similarly entail significant homopolar atomic pairings within the cell (below 16 K in the former, 48 K in the latter) which stand little betrayed by $a_o$ itself. For the A15 materials these changes do not break crystalline symmetry, as a result of the direct nature of the band overlap there, but for the case of the pnictides and FeSe a slight, symmetry-breaking, periodic lattice distortion almost ubiquitously shows up at lowered temperatures, encouraged by the soft layered form of their crystal structures. One senses this orthorhombicity to emerge as countering outlier to the potential superconductive instability. In the familiar CDW responses of 2H- and $1T\text{-}TaS_2$, at first sight hexagonal and trigonal like their host structures, there in fact occurs comparable slight loss of symmetry to orthorhombicity and triclinicity respectively [46]. Such symmetry breaking often arises in response to small additional constraints, such as the requirement for favourable *c*-axis stacking - not just of the CDW itself but also the entrained

PSD (periodic structural distortion) as well. In the current case there in addition is the question of magnetic coupling. We shall return to these sensitive matters in §§8/9.

The scenario envisaged then for the present superconductivity is as follows. The RVB condition ties up a great many of the spins in resonant non-magnetic pairings, promoted by the basic $Z=2$ unit cell with its strong fourfold n.n. Fe-Fe $\pi/\pi^*$ interactions. This pre-pairing of the bulk of the spins ultimately overrides the partnering drive towards a spin density wave. What the direct spin pairing accomplishes is to predispose the developing excitonic instability towards superconductivity, and, provided the potential lattice instability can be held off (often via substitutional or non-stoichiometric change) it becomes favoured option under the advancing correlation. The superconductivity finally achieved develops, as with the HTSC cuprates, in the highly favourable circumstance of pre-paired bosonic entities positioned degenerately with the residue of fermions. Together the two populations become incorporated into one overall superconducting state to build a $T_c$ much elevated in the highly resonant crossover conditions able to emerge under coordination unit shape ideality. Such a resonant boson-fermion crossover route to HTSC has a long history in the theoretical literature, e.g. Friedberg and Lee's papers of 1989 [47]. A considerably expanded exposition of this route to high $T_c$ very recently has been released by Byczuk and Vollhardt [48].

In the case of cuprate HTSC, I have over the years in an extended series of papers developed a detailed scenario based on such boson-fermion degeneracy, grounded in the mixed-valent situation prevailing there [49]. However with cuprate HTSC the bosonic pair is conceived as emerging essentially *within* a Cu-O coordination unit as local shell-closure $p^6 d^{10}$ entities, these generated as long-lived fluctuations within those coordination units being most driven towards trivalency by the highly localized substitutional and interstitial introduction of holes into the parent Mott insulators. I have introduced in order to describe such electron double-loading proximate to these sites the shorthand notation $^{10}Cu_{III}^{2-}$. The degeneracy of this bosonic object with the Fermi energy set by the remaining free carriers comes by virtue of the very marked negative-$U$ standing of that state, as the oxygen-$p$ and copper-$d$ states relocate strongly upon the termination of the $pd\sigma/\sigma^*$ interaction at shell closure. This electronic rearrangement and the ensuing induced superconductive outcome inevitably must within the narrow band system leave its mark in strong lattice changes, such as soft modes, unusual isotope effects, etc. However it remains in essence an electronically driven phenomenon. There is no retarded electron-boson mediated coupling here. The electron system itself is directly responsible. The above critical negative-$U$ situation is closely related to that which initiates disproportionation in say $CsAuCl_3$ or $BaBiO_3$,

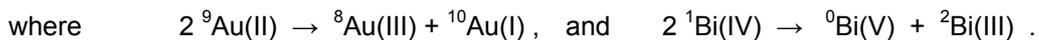
where  $2\ ^9Au(II) \rightarrow\ ^8Au(III) +\ ^{10}Au(I)$ ,  and   $2\ ^1Bi(IV) \rightarrow\ ^0Bi(V) +\ ^2Bi(III)$ .
These changes respectively involve $d^{10}$ and 'lone-pair' $s^2$ shell completion. For the mixed-valent setting of the doped cuprates such double-loading is however not frozen in, but proceeds via the dynamic boson-fermion degenerate resonance to enable the realization of HTSC and extensive pair-coupling. In the present case of the pnictides and FeSe the situation advocated is not quite so dramatic. The pre-formed pairs issue now not from double-loading essentially *on individual* coordination units, but from RVB-controlled spin-pairing *between* coordination units. In the prevailing semimetallic environment, the bosonic pairs of the bonding VB are able to transfer across into the degenerately resonant Fermi sea. The success of the overall pairing, original and induced, in establishing a superconducting outcome of note depends upon the fact that the small semimetallic overlap is indirect and hence subject to the further correlation-driven effects of the excitonic insulator process. The potential for the above procedure is accordingly as statistically remarkable as that being claimed for the cuprates. It stands just as sensitive in its detailed outcome to modification by applied pressure, chemical doping, structural perfection and distortion.

The slight distortions so frequently developing in the pnictides manage very effectively to hamper the above scenario. Not only do they break symmetry, but they facilitate too an entrée to magnetism and spin-pair collapse as the RVB coupling is disturbed. The rise and fall of $T_c$ under hydrostatic pressure, or chemical pressure, or substitutional and non-stoichiometric doping then become not so transparently related to the concomitant changes in coordination unit shape and bond splay angle $\alpha$, and the key role that this latter plays. For example with LaOFeAs under hydrostatic pressure, $T_c$ mounts not in fact from a drawing closer to shape ideality but because it brings elimination of the orthorhombic distortion. One is required to bear clearly in mind too the distinction in effect that hydrostatic pressure has here upon the 1111-system as against chemical pressure, from say substituting La by the smaller Nd. The primary effect of *hydrostatic* pressure on a *simple* layer compound like FeSe is disproportionately to cause a reduction in height of the softer vdW region of the structure. As the sandwiches concertina into this region, the process brings a decrease in $\alpha$. Of itself this should take FeSe to the *left* on figure 1, further away from shape ideality and toward lower $T_c$. Such a fall in $T_c$ would be precisely the opposite of what the above chemical substitution within the 'intercalated' layers secures through Nd substitution into La-1111, the basally constricting effect now driving the material on figure 1 *toward* shape ideality (lower $\alpha$) and to a higher $T_c$. That FeSe in fact experiences a *rise* in $T_c$ under pressure is very likely the consequence of its significantly augmented 3D band dispersion better facilitating inter-sandwich pair transfer. FeSe is in this respect structurally unique on account of its empty vdW region. 'Intercalated' LiFeAs, for which $\alpha$ likewise lies below ideal, and with a Mössbauer spectrum providing no sign at low temperatures of ordered magnetism or structural distortion, indeed exhibits the expected decline in $T_c$ under pressure [50].

At this point it will be a good test of the above general understanding to see if we can identify a dozen or so observations, hitherto often bringers of mystery and confusion to the literature, that now are able to receive ready clarification.

## §6. Shedding the mysteries.
### 6.1) Indications of resonant behaviour in the vicinity of $T_c$.

Several people have noticed that as one of these materials, whether under pressure or chemical stress, is taken closer to shape ideality and the optimization of $T_c$, the transition actually appears to get sharper.  This is odd as the inevitable inhomogeneity in a high pressure cell usually leads to the reverse effect.  It suggests the fine-tuning of a resonance effect.  It is accompanied by expansion in the range and intensity of the linear-in-$T$ resistivity found to show up universally in these materials directly above $T_c$ [26b,51].  Such a temperature dependence, which in the cuprates is very pronounced, has been taken in [49d] as evidence there of dominant boson-fermion scattering in the prevailing mixed, non-Fermi liquid condition. It is to be distinguished from that of the high temperature incoherence regime.

A good monitor of whether or not standard Fermi liquid behaviour stands behind the superconductivity developed in FeSe and the pnictides is the thermodynamic measure afforded by $\Delta C_p$, the jump in heat capacity across $T_c$.  In BCS mean-field theory $\Delta C_p/\gamma_n T_c$ (where $\gamma_n$ equals the normal state electronic specific heat) takes the constant value 1.43. With the present materials it pointedly has been commented that the discontinuity appears more variable, and a very recent paper from Bud'ko *et al* [52] shows the way to systematizing this.  They observe that over a wide range of materials in the 122 family the size of $\Delta C_p/T_c$ augments with $T_c$ strictly as $T_c^2$.  This constitutes a prime indicator once more of the resonant nature of the events being studied.  If $\Delta C_p$ were to adjust linearly in step with $T_c$, while at the same time $\gamma_n$ were actually to drop off linearly with the growth in $T_c$, then we would arrive at the above outcome.  The number of effective fermions appears diminished as $T_c$ climbs.

**6.2) Further evidence of a narrow normal state fermion pseudogap forming around $E_F$.**

The sign of the Seebeck coefficient (controlled by $-d\sigma/dE$) being positive in FeSe over the low temperature range somewhat above $T_c$, in which $\rho \propto T$, signals for this l.t. range a dominant *valence* band carrier number and/or mobility [53a,b].  As in SmFeAsO$_{0.85}$, once above this range, $S$ swings strongly negative to show by 110 K a *high* broad maximum, $\sim -50$ μV/K or more in Sm 1111-material [53c].  Such strong correlation effects also are on display in the significant temperature dependence of the Hall coefficient – see §7.  Additional direct evidence that due to strong correlations a form of deep pseudogap is present about $E_F$, over a considerable range above $T_c$, is provided by ultra-high resolution photoemission using a laser-excited spectrometer.  It supports a steadily deepening DOS dip below $\pm 10$ meV  [54].

**6.3) The shift in the Uemura plot.**

The striking thing about the Uemura plot (in addition to the fact that it gathers in all the exotic superconductors) is that its straight-line dependence of $T_c$ upon $n_s$ (the superfluid density) runs parallel to that for bosonic condensation. The plot finds itself displaced, however, significantly to the right (i.e. to larger $n_s$) as compared with the former.  This finding would imply that to attain a desired $T_c$, $n_s$ stands considerably higher than would be necessary were *every* pair of fermionic quasiparticles to constitute a fully effective boson. Viewed from the perspective of a given measured $n_s$ value, $T_c$ is down by a factor of about 5 on the associated $T_B$.  From this it is clear the *crucial* boson population in place is much below $n_s$.  Despite *all* the carriers ultimately becoming taken into the superconducting condensate, the majority are not those responsible for driving $T_c$ to such heights.  Such a circumstance was very evident in the data from YBa$_2$Cu$_3$O$_7$, where all the *chain* electrons are in the end taken into the condensate, but with virtually no change brought to $T_c$. The data which Uemura has to hand to generate these plots usually come from μSR precession signal relaxation rates measured in a magnetic field greater than $H_{c1}$ (i.e. in the vortex state). The very large signal decay rate that was found in [55] for LaOFeP as compared with La($O_{0.92}F_{0.08}$)FeAs marks how the quasiparticles in the phosphide are incapable of building a bosonic population of the same magnitude as that achieved with the arsenide. Note F-substitution in the latter material, by elimination of the orthorhombic distortion (and $T_N$), drives $T_c$ up there despite interferring with the $e/h$ balance. We earlier saw how the phosphide resides much farther from shape ideality. The impaired degree then of bosonic resonance produces not only a reduced fermion-boson interchange, but it encourages in addition the resurgence of magnetic fluctuations, changes which together result in a much-lowered $T_c$.

**6.4) A closer look at the pseudogap from NMR/NQR experimentation.**

Perhaps the clearest view of what is afoot here comes with the NMR results of Grafe *et al* [56], collected both for '$e$-doped' La($O_{0.9}F_{0.1}$)FeAs and pure LaOFeAs. (The 122 family [57] and FeSe [24c] insert certain complications better inspected separately, but all the data fall into the same general pattern.) The most striking feature is that the Knight shift, regardless of which atomic site type is being probed, senses a single spin-fluid and one that, as with the bulk susceptibility, is remarkably low in magnitude (and away from stoichiometry free from any trace of ordered, local moment behaviour). Furthermore $K_s$ is pointedly temperature dependent, settling to its minimum value only as $T \rightarrow 0$ K. Above $T_c$, $K_s$ mounts steadily very close to linearly with $T$. All this speaks of a correlation driven pseudogap, much as develops in the cuprates below $p = 0.185$. The pseudogap is non-states-conserving and distinctly unusual in that it does not lead to any weight-displaced peaks, just as much so for the Knight shift spectra as for the spin-lattice relaxation rate spectra. The absence of a Hebel-Slichter peak, along with the sharp fall off directly below $T_c$ in $1/T_1T$ vs. $T$ plots as $T^3$ or steeper (again as with the cuprates) has led to talk of d-wave superconductivity and nodes in the superconducting gap. The latter complex and important matter again is best dealt with subsequently via other more direct techniques. For the moment it suffices to concentrate upon the relaxation data above $T_c$. Here once again regardless of which nuclei in the unit cells one actually selects to sense the spin-lattice relaxation, there occurs little evidence of standard local moment formation, of their ordering, or of spin-fluctuation scattering therefrom. Conversely the relaxation data do not follow the simple Korringa form ($1/T_1T$ = const.) of a classical metal either. Instead $1/T_1T$ picks up precisely the same variance as did the Knight shift [56] in being $\propto T$, i.e. $\propto \chi_{spin}(0,0)$, i.e. $\propto$ the *fermionic* DOS – meaning the latter was being pseudogapped. That the various nuclei monitored each track the same temperature dependence means spin excitations have become suppressed uniformly over all $q$.

The sharp transfer shown above in $1/T_1T$ vs. $T$ at $T_c$, from variation above $T_c$ as $T$ to variation below $T_c$ as $T^3$ or greater, is not so readily evident with $BaFe_2As_2$ and FeSe, especially the former. Prior to reaching $T_c$, over a range of 100 K in $BaFe_2As_2$ [57] and of 75 K in FeSe [24d,58], one witnesses a significant growth in fluctuational scattering that appears not tied to any particular point in the Brillouin zone. The latter accordingly is *not* a reflection of the antiferromagnetic order present in these materials when at higher temperatures, in particular when undoped. Note the *Knight shift* actually stays virtually unperturbed right across $T_c$, but the above additional scattering and spin-lattice relaxation in fact terminates abruptly somewhat prior to $T_c$. The latter evidently has nothing to do with impurities or two-phase behaviour. Tellingly it becomes more marked, rather than less so, as the pressure on the above materials is increased [24d,59], tracking in this the rise there in $T_c$. The way I would suggest this low temperature scattering be understood is that it issues from the RVB/excitonic insulator/boson-fermion milieu itself. Under pressure, or in a more ionic setting such as that provided by $BaFe_2As_2$, the coupling of the lattice to the correlated, semimetallic system slows the fluctuational rate there to a level at which it becomes registered by the NMR probe, i.e. to $\omega < $ ~$10^7$-$10^8$ sec$^{-1}$. I see no justification to start turning here to standard antiferromagnetic spin fluctuations as responsible for the superconductivity, any more than I do with the cuprates, although the details of what is happening clearly are a matter to be pursued below.

**6.5) Information from inelastic neutron data on the 'resonance peaks' and spin gaps.**

Because people endlessly look to parallels with the cuprate HTSC materials, it frequently has been presumed that the unsubstituted pnictides are simple antiferromagnets – as if complementary to $LaCuO_4$ or $YBa_2Cu_3O_6$. However the magnetic ordering, met with in LaOFeAs at 140 K [60], is not into a classic Néel state, and nor indeed is it into a classic spin density wave state either. On the one hand it shows spin-wave excitations which are significantly gapped at low energy, and on the other it does not closely follow the Fermiology and take up an incommensurate **Q**-vector. Instead *inelastic* neutron scattering work reveals a strong, non-dispersing excitation at $(\pi,\pi)$ for temperatures in the range 250 K to 100 K, this scattering peaking quite sharply in intensity (over the above temperature range) at an energy transfer of around 8.5 meV. Note in all the pnictide families, the magnetic wave-vector extracted from the *elastic* magnetic Bragg peaks is similarly set at $(\pi,\pi,\pi)$, the corner of the primitive parent tetragonal zone – hence the source of the confusion.

As is to be seen from figure 9, in order to attain the observed magnetic cell would in a tight-binding picture rely on Fe-As-Fe superexchange, parallel in direction to the tetragonal cell edges and communicated via a pair of Fe-As (or Fe-Se) bonds subtending angle $\alpha$ [60]. The antiferromagnetic cell universally is observed in the company of a slight symmetry breakage to orthorhombic *Cmma* of about ½%, much as might be thought to arise from magnetostriction under the basal orientation of the spins. The latter are all set parallel and antiparallel to $b_O$, the slightly longer basal axis for the *orthorhombic* cell [60]. As tetrahedral shape ideality is approached under pressure or via the action of substitutional doping, it would appear that the above magnetic exchange steadily becomes superceded on cooling by direct spin coupling between nearest-neighbour Fe sites. These, recall, lie two to each tetragonal cell in the 45° directions, and even in Ce-1111 ($a_o$ = 4.00 Å) stand only 2.8 Å apart [61]. In figure 9 the basic tetragonal cell, observe, has been positioned in 'second setting' (see §3) to emphasize the equivalence of its two Fe atoms. RVB correlation-driven pair coupling would not necessarily demand any expansion here of the 'magnetic' cell, unlike the √2x√2 expansion actually beingwitnessed. Evidently things become considerably more intricate and interesting, as we shall find in §8.

Note with either orientation of the antiferromagnetic orthorhombic cell included in figure 9, the two Fe atoms within each *individual* tetragonal cell are everywhere spin-opposed. This fact serves to emphasize the ease with which these materials are free to slip from the long - range order (LRO) site-moment organisation of 'high' temperature antiferromagnetism over to SRO bond-pairing or to RVB as correlation conditions become appropriate down towards $T_c$.

While for systems near shape ideality the spin susceptibility on cooling becomes basally isotropic and experiences a steady decline, such that in due course the systems become spin-gapped and effectively non-magnetic, this does not imply FeS, say, will have become *on-site* S=0 like $FeS_2$. Rather we have a spin pseudogap condition resembling the one in the cuprate HTSC materials. As $T$ in LaOFeAs drops below a potential low $T_c$ this spin gapping extends up rapidly to higher energy to expose by 7 K a sharp 'resonance peak' at 12 meV [62]. In the case of '*e*-doped' $Ba(Fe_{1.85}Co_{0.15})As_2$ the spin gap (as measured on compressed powder) is found fully openned initially only to 3 meV, but once well below $T_c$ (26 K) the final resonance peak in evidence (formed again at the π,π equivalent) materializes up at 14 meV [63]. Taking the latter feature to identify $2\Delta(0)$, the energy converts here to the equivalent of $4.3kT_c$. With reference to the classical mean-field (*s*-wave) value for the ratio $2\Delta(0)/kT_c$ of 3.54 the above figure would mark a strong-coupling situation.

The reason why, within the present modelling, one can associate the above resonance peak energy with $2\Delta(0)$ is that it is perceived as being yielded by the RVB bosonic spin-pairs, and in the global superconducting state these pairs have stabilized to lie below $E_F$ by binding energy $U$ per electron. To generate two independent fermions from such an S=0 bosonic entity demands essentially an S=1 spin flip, precisely as is secured in inelastic neutron spin-flip scattering. The difference between the present case and what is understood as occurring in the HTSC cuprates is that the boson in question now has been sourced from within a couple of coordination units, whereas in the cuprates it was effectively sourced by just the one such unit presenting $p^6d^{10}$ shell-filling [49c]. In the cuprate case the extent of the mixing and coupling between the bosons and the degenerate fermions proves in consequence the greater and $T_c$ the higher. The gap function $\Delta(T)$ measured in the HTSC cuprates emerges even less classical in form than what is seen here with the pnictides.

It will be good to have single crystal data from the various pnictides in due course to sharpen up these observations and to permit a more precise and meaningful comparison between the different systems over the matter of $\alpha$-ideality. Qualitatively however the association of the resonance peak, strongly in evidence once below $T_c$, with the dominant pair binding energy looks well founded. A very recent paper from Zhao *et al* [64] dealing with $Ba(Fe_{1.9}Ni_{0.1})As_2$ ($T_c$ = 20 K) has just reported that in a strong magnetic field the size of both $T_c$ and the intensity and energy of the resonance indeed become simultaneously suppressed.

**6.6) The bosonic mode as registered by optical infra-red (IR) spectroscopy.**

Timusk and coworkers following their ground-breaking work upon the cuprates [65,66] have now for a pnictide superconductor obtained a comparable extraction of the boson coupling function $\alpha^2.F(\omega)$ from the measured optical scattering rates, $1/\tau(\omega,T)$, following a maximum entropy deconvolution routine [67]. IR reflectance measurements were made on a single crystal of $(Ba_{0.55}K_{0.45})Fe_2As_2$ with $T_c$ = 26 K, and expose changes to the spectrum arising below 300 K over the entire energy range under 150 meV. Below 100 K new features peak up strongly there into a doublet at 10 and 30 meV followed by a broader hump centred at 90 meV (720 cm$^{-1}$). Although the pseudogap forming in $1/\tau(\omega)$ falls in the phononic range, the measured temperature dependence and spectral details are not consistent with a phononic origin. Remember the system is a metal and phonons then figure only weakly in the spectrum. Boeri, Dolgov and Golubov [39] have calculated the coupling parameter ascribable to phonons to be no more than $\lambda_{ph}$ = 0.21, whereas the current experiments and analysis uncover $\lambda$ values already ten times greater than this by 80 K, and growing to 3.42 by $T_c$. As was stated above, we are looking below $T_c$ to a value of the energy for the bosonic resonance of about 15-20 meV. In practice to extract the actual boson energy from the self-energy related reflectance traces calls for detailed analysis, as has more recently been carried through with the HTSC cuprates in [66]. With the latter materials the boson energy itself emerges as being roughly 40% smaller than the energy of the associated reflectance feature, and accordingly now the strongest spectral feature, standing at just under 30 meV, would fit current expectations. In bosonically induced global superconductivity note that $T_c$ and $2\Delta(0)$ always are going to reflect the weakest link in the overall move to condensation.

The uppermost anomalous feature around 90 meV, being sited toward the lower end of the range where $\rho \propto \omega$ holds, looks in contrast to be connected with the changes bestowing incoherent behaviour; changes that for the cuprates were claimed to incorporate strong lattice relaxation [49d].

**6.7) Injections from ARPES.**

The ARPES results are forthcoming mainly from the Ba-122 family because of the availability of suitable crystals. They fall into several strands and appear sufficiently in line with expectation to generalize. As with the cuprates they support the above discussion in manifesting the presence of a kink in the low energy dispersion curves [68], indicating thereby the action of the intersecting bosonic state in imparting strong renormalization to the effective masses of the carriers around $E_F$. The reduced overall width of the V.B being reported expresses likewise the appreciable correlation beyond that carried by the LDA exchange approximation. The semimetallic pockets as disclosed by ARPES, although somewhat smaller in size than those calculated, do, however, remain in the anticipated locations in the B.Z.. A more striking observation is that once below $T_c$ the superconducting gaps on those pockets are shown not to be equal in size – although seemingly roughly isotropic in direction. In this they are reminiscent of what was found with $MgB_2$ (semimetallic between B-B $p_{x,y}\sigma$ and $p_z\pi$ bonding and antibonding states, and exhibiting a $T_c$ of 40 K) [69]. Again now the largest of the gaps arises in a hole band and is located about $\Gamma$.

Drawn from the more recent literature, reference [70] undertakes for Ba-122, pure and substituted, a detailed ARPES study above $T_c$ of the LDA band shifts and the mass renormalizations required to agree with the X-point data ($\equiv \pi,\pi$ in the basic tetragonal zone), taken from its two electron pockets. The results display surprisingly little sensitivity as to whether the sample in fact is h-doped ($K_{0.4}$) or e-doped ($Co_{0.06}$). Such ARPES data are in line with the more detailed numerical assessments acquired from the dHVA/dHS experiments. (The latter are pursued, recall, within the 'normal' state at very low temperatures after suppression of the full superconducting condition by the application of magnetic fields in excess of $H_{c2}$, or rather $H_{irr}$, i.e. > 30 tesla [29]). As stated back in §4, the overall effect of the electron correlation active near $E_F$ is to deepen the pseudogapping there via a shifting of both sets of bounding bands by ~50 meV, the h-bands around $\Gamma$ being moved downwards, the e-bands around $\pi,\pi$ moved upwards.

As regards the situation *below* $T_c$ and direct observation of the superconducting gap sizes, recent ARPES work from Xu *et al* [71] comes for a $K_{0.4}$ sample ($T_c$ = 37 K) to the following estimates $\Delta^\Gamma_\alpha$ = 8.5 meV, $\Delta^\Gamma_\beta$ = 4.0 meV, and $\Delta^X_{\gamma,\delta}$ = 7.8 meV. These values drop in their less-substituted sample ($K_{0.25}$, with $T_c$ = 26 K) by about 30%. Such a reduction in gap size with fall in $T_c$ (appropriately found too to scale with $H_{c2}$) is very different from the form of pseudogap behaviour witnessed in the underdoped cuprates. Although a pseudogap is forthcoming in the pnictides it is not associated with the 'dichotomy' in gap sizes witnessed with the cuprates. There, what is more, the behaviour around the Fermi surface was highly anisotropic under an order parameter of $d_{x2-y2}$ form. Now in the pnictides this most certainly is not the case. The latter matter has been probed directly in the scanning SQUID microscopy work of Hicks *et al* on F-substituted polycrystalline Nd-1111 [72], and no $\pi$ phase-shifts are in evidence, no orbital currents being observed. Such a finding does not mean the order parameter here is of simple s-wave form. As with $MgB_2$ the order parameter actually would appear to be extended s-wave ($s^\pm$: $\Delta(\mathbf{k}) = \Delta(\cos k_x + \cos k_y)/2$ ), this entailing a sign change in $\Delta$ between the hole pockets at $\Gamma$ and the electron pockets at $\pi,\pi$. Such an order parameter actually introduces no nodal behaviour into the current semimetallic Fermi surface. Good reason exists theoretically to anticipate the above outcome [73]. However again we shall find below how things turn out somewhat more complicated in practice.

**6.8) Revealing input from electronic specific heat work.**

It is possible the two-peak behaviour extracted from electronic specific heat data for FeSe [33] is making record of the above two-gap state, much as previously has been recorded in the cases of $MgB_2$ [74] and $NbSe_2$ [75], and similar attention should now be paid to LiFeAs ($T_c$ = 20 K). For the latter present indications are that the jump in $\Delta C_p^{el}/\gamma_n T_c$ across $T_c$ is considerably below the BCS value of 1.43 [76]. By contrast Mu *et al* [77] working with $(Ba_{0.6}K_{0.4})Fe_2As_2$, where $T_c$=36.5 K, emerge with a $\Delta C_p^{el}/\gamma_n T_c$ value there of 1.55. This now is somewhat greater than the mean-field value, and it arrives in conjunction with a very sizeable value extracted for $\gamma_n$ of 63.3 mJ/mol K$^2$. The latter proves an order of magnitude greater than was obtained with F-doped LaOFeAs [78]. The likewise high value of the $\Delta C_p^{el}$ jump at $T_c$ in the 122 material means that the feature stands out clearly there in the raw data. The empirical value of $\Delta C_p^{el}/T_c$ is up very close now to 100 mJ/mol K. Such magnitudes both for $\gamma_n$ and $\Delta C_p^{el}/T_c$ (remember we are dealing with a semimetal) speak of strong correlation and coupling. For a simple metal $\gamma_n = \tfrac{2}{3}.\pi^2.k_B^2.N(E_F).(1+\lambda)$. In the present case one may accordingly attribute the above results to elevated values for both $N(E_F)$ and $\lambda$ over those expected under electron-phonon coupling, arriving from the action associated with the pseudo-gap, including raised values for $m^*$, and they match our electronically-based understanding of what drives the pnictide superconductivity. The level of differentiation observed between one pnictide material and another in these matters comes, as previously was indicated, from the degree of proximity to shape ideality. Thus the $\Delta C_p^{el}/T_c$ jump observed in LaOFeP-based material [79] stands well below that seen even with LaOFeAs.

Recently Bud'ko, Ni and Canfield [52] have arrived at the very telling observation, when examining a whole series of BaFe$_2$As$_2$-sourced materials, that $\Delta C_p^{el}/T_c \propto T_c^2$, this relation involving such a large coefficient of proportionality that virtually two decades in $\Delta C_p^{el}/T_c$ are spanned here. Because with standard superconductors $\Delta C_p^{el}/\gamma_n T_c$ = a constant, behaviour like this would intimate linear augmentations as $T_c$ both to $N(E_F)$ and $\lambda$, as shape ideality and the resonant superconducting condition are approached.

Given the extent and level of the analysis which Mu et al [77] have undertaken on their $C_p^{el}(T,H)$ data, it is rather surprising they did not attempt a multigap treatment in light of the clear manifestation of a secondary hump within their $C_p(T)$ trace around 20K, much as was exhibited by MgB$_2$ and NbSe$_2$. Already we have observed from the ARPES work that the order parameter for pnictide systems when approaching shape ideality is not simple s-wave, but multigap $s^{\pm}$, wherein different levels of gapping are supported across the semi-metallic k-divide. Accordingly the gap value of 6 meV extracted by Mu et al and subsequently inserted into their further analysis must stand as average value over the span of 'local' values, such as those cited in the previous section deriving from the ARPES data. Ma et al do however very elegantly and effectively probe the magnetic field dependence of the specific heat data to confirm that their BaFe$_2$As$_2$-derived material displays node-free superconductive behaviour. This result once more is in striking contrast with their earlier results on F-substituted LaOFeAs material, for which the analogous data would clearly indicate the occurrence in that case of nodes in the superconducting gap [78]. This has not simply to be taken as marking a distinction in behaviour between the 122 and 1111 families: F-substituted SmOFeAs is known to show node-*free* behaviour [79]. It is that matter of shape ideality again (see fig.1).

**6.9) Comparable results from optical measures of the electronic kinetic energy.**

It is very striking that the Drude IR edge in the high $T_c$ pnictides stands clearly displaced towards low frequency as compared with what LDA band structures would lead one to anticipate. Qazilbash *et al* [80] very recently have made a close assessment of this correlation-driven shift as a function of temperature both for LaOFeP and LaOFeAs. By conversion of the measured optical conductivity, $\sigma_1(\omega)$, through to the associated optical scattering rate, $1/\tau(\omega)$, it immediately becomes apparent that, as with the HTSC cuprates [81], the correlations run at such a level as to drive the carriers in these semimetals close to incoherence. $h/\tau(\omega)$ falls only just short of $kT$, and in fact climbs linearly with $T$, pursuant to the linear rise with $T$ in evidence in the d.c. resistivity just above $T_c$ [51]. Such behaviour is very like that recently re-examined by Cooper *et al* [82] for the HTSC cuprates, and which I interpret as resulting from boson-fermion scattering in the resonant negative-$U$ crossover circumstance prevailing there [49e]. In the cuprates the high degree of correlation in evidence reflects their proximity to the Mott transition. Now in the pnictides however it is sourced by proximity to shape ideality and the associated deep semimetallic pseudogapping. The resonant bosons are this time of RVB origin and experiencing excitonic insulator coupling to the fermions. Because LaOFeAs stands much closer to shape ideality than does LaOFeP, such correlations, and hence $T_c$, are greater in the *former*. This marks a striking inversion of fortunes as compared with Mott-driven correlations, where a phosphide always will exhibit stronger correlations than does the corresponding arsenide, due to the more advanced *p-d* mixing in the latter – witness CrP vs CrAs, NiP vs NiAs (see [2a fig 1]). Qazilbash *et al* [80] numerically assess the situation via appraisal of the optical conductivity through the integral -

$$hc_o/_{e2}.\int_0^{\omega_c} 2h/_{\pi}.\sigma_1(\omega).d\omega$$ , this being equivalent to the electronic kinetic energy $K_{exp}(\omega_c)$.

An appropriate frequency cut-off of 3,000 cm$^{-1}$ has been applied here. The authors make comparison then of this $K_{exp}$ with direct evaluation of the quantity $K_{th}$ drawn from the much more lightly correlated LDA band structure. The experimental value for $K$ proves diminished relative to the latter by a factor of 2 to 4 times – and the arsenide indeed stands here the *more* correlated of the two. In [80] this was presented as a mystery.

**6.10) The penetration depth results**.

In like vein to the above, we in Bristol initially were perplexed by the conflicting outcome to our penetration depth measurements, made first on Sm(O/F)FeAs [83] and then subsequently on La(O/F)FeP [30]. In line with the above it gradually became clear why the former system yielded the exponential variation of a fully-gapped response for $\lambda(T)$, whilst the second system exhibited a low temperature power law variation for $\lambda(T)$, in keeping with a partially-gapped status. The node-free behaviour of the $s^{\pm}$ order parameter is secured with the shape-ideal, spin-paired, small overlap Sm system, whilst a node-bearing response, materializing under order parameter $\Delta = \cos k_x/2.\cos k_y/2$ , is adopted with the system well-removed from shape ideality, in which magnetic input is not being fully subsumed into RVB and the deep spin-singlet pseudogapping to follow. Both FeSe [84] and also BaFe$_2$As$_2$ and derivative material [85], which, when unsubstituted, stand at some distance from shape ideality and RVB behaviour, do not display a truly exponential variance in $\lambda(T)$, but show at low $T$ a $T^2$ dependence, as for dirty $s$-wave. The 'dirt' now however is intrinsic, with the hole and electron pockets behaving differently.

Although the two above order parameters each are covered by representation $A_{1g}$ within point group $D_{4h}$, their geometrical arrangement of nodes is so different that it is hard to see how the situation encountered can slide continuously from the one to the other without some first-order break being encountered, especially when folding in the changing role of the lattice. We shall return in §8 to examine the light recent experiment casts on this matter – once more a situation clearly more complex than was originally suspected. Prior to looking into this, some further comments are first in order on the role of chemical substitution.

### §7. A further look at chemical substitution.

The 'doping' terminology so widely used with the pnictides is, as with the cuprates, one that does not serve the field well. It is one that with the cuprates I sought from the start to avoid, advocating use of the terms 'substituted' and 'interstitial' [49a,b]. The reference point for an HTSC cuprate is to the $d^9$ Mott-insulating parent and the half-filled $d_{x^2-y^2}$ band. By the time metallicity is met with, however, the number of active carriers is rapidly departing from the number of monovalent 'dopant' centres introduced. In $(La_{2-x}Sr_x)CuO_4$, etc. '$p$' in reality monitors the Cu(III) count, not the carrier count. Now, too, Co substitution in say $Ba(Fe/Co)_2As_2$ does not simply see electrons added to the conduction band, but also fills in holes in the valence band of the parent compound – this time no Mott insulator but a band semimetal. The false parallel being imposed on the pnictides has come from a desire to endow the present antiferromagnetic parents with comparable characteristics to the Mott-insulating cuprate parents like $La_2CuO_4$, and then to drop into attributing the superconductivity to spin fluctuations. It is an attribution made too lightly, and one which patently is as false now as then.

In both fields $T_c$ climbs as low energy spin gapping grows. With the pnictides the latter is fostered by the move to shape ideality within the tetrahedral coordination unit, with a favouring of local moment-freed RVB spin pairing, as opposed to antiferromagnetism – even if more related to SDW-type nesting, without the Hund's rule type behaviour seen in hexagonal FeS. In the cuprates there comparably is in play a shape-dependent aspect to the rise of $T_c$, expressed there in the degree of distortion away from ideality of the octahedral Cu-O coordination unit in the form of a strong Jahn-Teller type $c$-axis elongation. This distortion is greatest in the Hg-cuprates for which $T_c$ comes to its maximum. With the cuprates the primary effect is to diminish the basal lattice parameters and thereby hold down magnetic moment formation, extending the spin gap to its maximum energy – and specifically beyond the S=1 spin-flip excitation energy of the sustained superconductive pairs. The precise aim in selecting the ideal substituent with the cuprates is to secure the lowered basal lattice parameter and full suppression of free moments precisely as the key concentration of Cu(III) centres of $p$=0.16 is reached. The latter is a percolation/charge-stripe dictated level [49,86]. With the pnictides it is evident from experiment that likewise there exists a comparable drive to offset maximal $T_c$ from the simple stoichiometry of the parent. This action now though is not to generate metallicity or to acquire some specific level of charge deviation, but rather to perturb the system sufficiently that the excitonic coupling across the semimetallic divide then favours superconductivity as alternative to the vying instabilities of SDW and CDW/PLD. Here disordering is clearly more quickly advanced by substituting directly on the Fe site [87] than it is when disturbing the electron/hole balance at one remove, as in substituting Ba with K [88]. Breaking the regularity of the site up-spin/down-spin *d*-electron array automatically will favour RVB over well-structured antiferromagnetic order. The current $Z$=2 cell geometry helps greatly in this regard, as too does the dominantly 2D-coupling of the layer compound environment.

It is highly informative to consider precisely what Co substitution into $BaFe_2As_2$ [87] is bringing about. Just 8% Fe replacement leads to a complete elimination of the parent SDW ($T_{onset}$ = 135 K), the maximal $T_c$ of 23 K then immediately being acquired. This substitution in fact is associated with a reduction in average lattice parameter [89], as if resulting from low-spin $d^6$ Co(III). The surplus electrons then must be delocalized over the Fe sites. That such disturbance promotes low temperature spin-pseudogapping is registered both in $^{75}$As and $^{59}$Co NMR work [90]. No local moments are induced here, and the Knight shift data reveals a strong reduction in spin susceptibility with increase in $x_{Co}$ – as with decrease in $T$ and right from room temperature. By $x_{Co}$ = 0.1, $K_s$ has levelled out at a low and temperature-independent value, with the same being witnessed for $1/T_1T$. Both spin susceptibility and spin-fluctuation-induced relaxation rate have become invariant and small in the pre-superconductive range (as earlier seen in §6.4 with the 1111-materials).

The present research from Ning *et al* [90] proceeds further and discloses, from study of the NMR lineshapes, that the changes are apparently associated with micro-inhomogeneity at the nanometre level, here the substitutive changes very locally controlling the detailed response. Such disturbance led within low-*p* cuprates to the development of striped behaviour, both fluctuating and frozen, but in the pnictides there as yet is no sign of comparable micro-organization. There are, however, indications of some macrostructural adjustment as the magnetic organization is disrupted and RVB takes over, as we shall see below. Ning *et al* in their very interesting contour plots of the NMR data, drawn up across the $T$–$x_{Co}$ plane, reveal $1/T_1T$ and $K_s$ to have dropped respectively to 0.28 and to 0.25% by the point at which $T_c$ globally maximizes (23 K, here at $x_{Co}$ = 0.08). Their plot has been reproduced with some amplification as figure 10. Many, pursuing the spin fluctuation scenario for the superconductivity, would like to point to such a figure as indication of magnetic, quantum critical behaviour, with the latter somehow associated with seeding-out superconductivity. My own reading of the above figure is that in pure $BaFe_2As_2$ superconductivity would have arisen near 40K were it not for the preferred adoption of the SDW state onsetting at 138K. The two states are seen as competitors – not working in tandem. The spin pseudo-gap energies as deduced from this NMR work amount to around 500 K or 40 meV.

One might ask on what experimental grounds such a viewpoint, such an extrapolation as the above, is based. I would like to point here to the high pressure results secured by Kotegawa *et al* [91] from *un*substituted $SrFe_2As_2$. The latter enters a SDW state below 200 K for which the onset temperature is found to decrease rather little as a function of pressure. Then suddenly between 3 and 3½ Gpa it rapidly declines directly to be replaced by superconductivity with a $T_c$ of 34 K. $T_c(P)$ above this pressure would again look to extrapolate back to ~40 K at $P$ =0. Given the inhomogeneity within the pressure cell, this data strongly suggest a first-order replacement of the one ground state by the other. What has swung the balance now in favour of superconductivity? It is known that pressure deepens the spin pseudogap, just as did Co substitution and the lattice parameter decrease which it introduced. A further comparable and striking result is obtained when working in the *isovalently* 'anion'-substituted system $BaFe_2(As_{1-x}P_x)_2$ [92]. Despite this replacement of the arsenic by phosphorus actually forcing the system further away from shape ideality, the disorder that it introduces and the reduction in basal parameter are sufficient for a while to favour superconductivity as ground state across the substantial range from $P_{0.30}$ to $P_{0.65}$. From there once again $T_c(x_P)$ would extrapolate back under the SDW state of pure $BaFe_2As_2$ to a value of 40 K. In a small range around $x_P$ = 0.3 the two states appear to coexist (probably micro-inhomogeneously – see below). $BaFe_2As_2$, note, does itself in fact become superconducting under 4 Gpa, but by then with a $T_c$ of only 29 K [8].

Being layer compounds the pnictide materials can rather easily be perturbed in the third dimension, and that probably is how K substitution of Ba in 122-material ought in large measure to be regarded. However the charge change simultaneously introduced makes such attribution there rather tenuous. A very striking new result has, however, just been secured pointing in that direction. Intercalation of $SrFe_2As_2$ by water vapour has been discovered to bring about superconductivity with $T_c$ at 25 K [93]. There is no charge change here, just perturbation of the structure and of the Fe-based 3D spin ordering. A rather similar result was experienced with trigonal prismatically coordinated $TaS_2$, where intercalation by a wide variety of organic molecules suppresses the CDW there in favour of superconductivity.

Above in §2 we have noted that in LaOFeP the LaO there plays the role of a monovalent lanthanyl ion. The La atom itself here is on the *out*side of the LaO sandwich and feeds its third electron over directly into the P atom. The latter atom hence forms an integral part of the La atom's coordination shell. In a great many La compounds its coordination shell is expanded beyond 6; for example it is 8 in the $d^1$ layer metal $LaI_2$ [94]. What further pseudo-ions might now be intercalated to play the same role here of supplying electrons to the FeP sandwiches, and be able to prise the latter even further apart? The answer, steadily building, is a great many. A particularly striking one is $Sr_2ScO_3$, or rather a block twice that thickness, the FeP layers now becoming spaced by in excess of 15 Å. This Sc(III) phosphide product is *non*-magnetic and is superconducting with a $T_c$ of 17 K [95]. In marked contrast the corresponding arsenide is this time non-superconducting, in fact non-metallic, and it exhibits Curie-Weiss behaviour at the Fe sites [96]. Such inversion here between arsenide and phosphide looks to be due to the lifting of the semimetallic overlap – see [96a, figs. 3 and 4] and [96b fig 4] for the effect on $\rho$, $\chi$, $S$ and $R_H$. Then with $(Sr_2\underline{V}O_3)_2.Fe_2As_2$ and reduction again to the $a_o$ parameter (and $\alpha$), a remarkable re-emergence of $T_c$ at 37 K is witnessed [97]. Note here that, as with LaO, both above inserts inject into the FeX sandwich one electron per intercalate molecule, procuring thereby our basic 'stoichiometric' $d^6$Fe(II) electron count. Analogous results are now forthcoming by insertion of various complex titanates [98]. Whether or not all these intercalates in fact stand commensurate with the host is yet to be determined. Incommensurateness would add further to the disordering process. 'Misfit intercalation' is well-known for layer systems, $(BiS)_{1.11}.NbS_2$ being a prime example [99].

The complete break-through above in $(Sr_2ScO_3)_2.Fe_2As_2$ of a semiconducting gap at $d^6$ arrives within a material that remains tetragonal at low $T$. There is even greater danger of this occurring whenever $e$-$h$ nesting drives the system into a low-temperature density-wave state, cutting down further the residual Fermi surface. Several weakly substituted, high $e/h$ balance, systems have actually been witnessed, through study of their Hall coefficients, to suffer 'metal/insulator' conversion below $T_{DW}$, in particular when they stand close to high shape ideality and already contain a well-developed DOS pseudogap; Sm, Ce and even La($O_{1-x}F_x$).FeAs [6a,100,101] are cases in point, as well as $BaFe_2As_2$ [102]. A close study of the Hall coefficient in the Sm-1111 system, made both in the high- and low-field regimes and as a function of fluorine substitution, has demonstrated how at low $x$ the number of free carriers declines exponentially once below $T_{DW}$ [103]. No carriers of course means no superconductivity. Sm-1111 with $x = 0.12$ sees $n$ become $< 10^{19}$ cm$^{-3}$ by 4 K. A final elimination of the DW state at $x = 0.18$ frees the potential in the system to manifest superconductivity immediately with $T_c^{max} = 54$ K. The fact that in $(Sr_2VO_3)_2.Fe_2As_2$ $T_c$ climbs yet further from 37 K to 46 K under pressure [104] would similarly indicate that a certain DOS level and metallicity is best suited to uphold the excitonic superconducting state.

### §8. The engagement of the lattice.

Above we have seen some of the effects wrought by chemical substitution upon the lattice and $T_c$, and at several stages have noted earlier the sensitive response of $T_c$ to pressure. A further striking observation related to the lattice, very recently reported in [105] is the change to $T_c$ in strained (unrelaxed) epitaxial thin films generated by altering the lattice mismatch to the substrate. Using substrates of $YAlO_3$, $LaAlO_3$, (La,Sr)(Al,Ta)$O_3$ and $SrTiO_3$ (of increasingly larger $a_o$) it has for the case of $Ba(Fe_{0.9}Co_{0.1})_2As_2$ been found possible to induce a linear rise of several degrees in $T_c$ through altering the $c/a$ ratio conferred upon the film. From the lattice parameter details supplied it is apparent that $\alpha$ rises and $T_c$ falls with YAO as substrate, whilst with LSAT and STO $\alpha$ falls and $T_c$ rises. LAO affords virtually a perfect lattice match for the Co-substituted 122 material (where $\alpha$ is somewhat above ideal) and $T_c$ remains unchanged. Such effects inevitably imply there are going to be isotope effects in these systems – and there are [106] – but that does not mean $T_c$ is primarily controlled by electron-phonon coupling. In any reasonably narrow-band circumstance it becomes impossible to decouple charge or indeed magnetic effects from the lattice.

At $T_{SDW}$ the measured thermal expansion coefficients $\alpha(T)$ show something like a 20% drop [107], a change that in fact follows in quick succession, for the 1111-LnO family, upon a comparable decrease at a sharply defined structural distortion temperature occurring some 5 K to 25 K above $T_{SDW}$ [107]. I would like to read this overlying structural transition as a CDW. Certainly it is not a consequence of magneto-striction because it clearly preceeds the magnetic order (and indeed outlives it). However, at the same time, it presents the same essentially √2x√2 increase in basal cell geometry (see fig. 9). In this for 1111, as with all related families, the increase involves transferring from the 4Å-edged primitive tetragonal cell to a face-centred cell of orthorhombic form for which $a \approx b$ ~ 5.6 Å.. In all cases the level of symmetry breakage is slight {$o = (b-a)/½(b+a)$ ~ 0.5%}, although the transition itself can manifest a small first-order component [108 fig.3]. With *un*substituted 122-material, the magnetic and the distorted conditions are (untypically) entered into simultaneously, and this has the profound effect of changing the spin-fluctuational behaviour monitored by neutron scattering from being 3D-Ising-like to being 2D-Ising-like [109]. The intimate effect of the slight structural change upon the magnetism is echoed by an equally striking effect of the lattice distortion in regard to the subsequent superconductivity, as evidenced in the isotope effect. Whether with (Ba/K)Fe$_2$As$_2$ or Sm(O/F)FeAs different phonons have been observed to experience different isotope shifts, but in each individual case both $T_c$ and $T_D$ are found remarkably to show identical shift exponents [106a].

However, as we have witnessed above, where $T_c$ is highest, as in Sm-1111 or in FeSe under P, the Mössbauer data makes very evident that the ordered antiferromagnetism and much of the paramagnetism have by then been shed. It is clear, what is more, neither magnetic order nor the orthorhombic lattice distortion (LD) are obligatory forerunners to the emergence of pnictide superconductivity, each being absent in LiFeAs [6a] and Sr$_2$ScO$_3$.FeP. In the case of FeSe the (initial) orthorhombically induced quadrupole splitting of the Mossbauer 'singlet', moreover, becomes lost under pressure *as $T_c$ rises* [26b fig2b]. Such sensitive interplay and competition between these three aspects, SDW, CDW/PLD, and superconductivity, would suggest all emerge from the same origin. All three are able to stand as expressions of the excitonic insulator interaction, and extremely small free energy changes are implicated in tipping the balance between the possible ground states. These ground states may involve a coexistence of SDW and/or CDW/PLD with the superconductivity, or the superconductivity can exist alone. In fact where $T_c$ is highest the latter circumstance would look to hold. Where these 'manifestations' coexist, this either may be accomplished homogeneously (as witnessed previously in 2H-NbSe$_2$, with the presence below 6 K of both CDW/PLD and superconductivity), or it may involve heterogeneous phase separation into micro-regions, the latter variously dominated by one or another aspect. A rather similar situation was encountered in the HTSC cuprates in the case of La$_2$CuO$_{4+\delta}$ under certain interstitial oxygen overloads, $\delta$. There, as was revealed by use of dark-field electron microscopy, some regions are antiferromagnetic, and others superconducting, dependent upon the local organization of the interstitial content [110].

In the present √2x√2 structural transitions little or no discontinuity in underlying basic cell volume is discerned, but there is a sharp change, as pointed out above, in thermal expansivity, the latter becoming somewhat reduced below the transition (see [108] fig.3). Where perhaps more revelatory effects are detected is (as with $MgB_2$) to specific phonons. As previously has been indicated, some exhibit an isotope effect whilst others do not [106a]. LeTacon *et al* [111] working with small Sm(O/F).FeAs crystals and employing inelastic X-ray scattering very recently have reported *oppositely-signed*, doping-induced, renormalizations for phonons having energies lying close to either side of a potential resonance at 23 meV. It is too early as yet to come to firm conclusions here, since LeTacon *et al*'s results are for 300 K only. It will be most interesting to pursue what developments arise through $T_{CDW}$/$T_{SDW}$. Anyone who doubts the marked response of the lattice to what is occurring electronically out well in advance of $T_{CDW}$/$T_{SDW}$ should look at the recent resonant ultrasound spectroscopy results from Fernandes *et al* [112] revealing very strong softening of the shear modulus.

Figure 11a represents the type of CDW/PLD √2x√2 supercell array compatible with the presence of RVB coupling and able initially to uphold tetragonal symmetry within the present unusual network of edge-sharing tetrahedra. The original primitive tetragonal Fe-X units are marked out in 'second setting' and the √2x√2, 45°-rotated, (tetragonal) supercell shown dashed. At the latter's centre lies a resonating square plaquette of four spins. If these units were to relax somewhat, one ought to be able to pick up appropriate superspotting by electron microscopy if not by X-ray and neutron diffraction. Unfortunately in many of these materials this will not be easy. To date there is definite indication of additional diffuse scattering being present above the orthorhombic symmetry-breaking distortion temperature, and we shall look at this matter in more detail in §9. Whether under instigation from non-stoichiometric *e*/*h* balance, or from substitutional disorder, or from *c*-axis stacking adjustment, figure 11b illustrates how the above tetragonal array of bonds is able readily to give place to orthorhombic distortion of the √2x√2 supercell simply by this freezing out of 1D pair-striped domains, these latter taking up twinned orientation as portrayed in figure 11 b. Alternatively, given the propensity of the systems for face-centring, one might well encounter the 1D arrangement of dimer pair bonding portrayed in figure 11c, again in conjunction with its twin.

### §9. Microstructural behaviour in FeSe and the pnictides, and what it is able to reveal.

The microstructure of twin formation arises from local relief of lattice strain as the new state breaks lattice symmetry. The faster the temperature decrease and the larger the impurity/structural defect density to initiate twin boundary pinning, the smaller will be the typical twin domain size. This fracturing of the crystalline continuity is to be distinguished carefully from inhomogeneous two-phase behaviour, both as to its origins and form. The latter process follows breakdown in solid solution formation between components A and B over some concentration span. In the thermodynamics controlling solid solution, free energy lowering is in general enabled through the downward bowing of the configurational entropy term, but this form can be disrupted by the local A-B interaction becoming repulsive over a certain concentration range (or by the descent of new ordered states). Such repulsion can result from a simple size effect mismatch between components or more interestingly be electronically based, as when trying to mix a Mott insulator with a band material. If the above changes become sufficiently pronounced as to produce an actual upward dimple in the free energy curve then the system finds it energetically advantageous to break up into a two-phase mix, with individual compositions and their relative amounts set by the 'common tangent' construction. The latter equates $\partial G/\partial n$ at the two tangential points, defining thereby state equilibrium (see [113] for the physics behind the construction of phase diagrams). Sometimes component A will tolerate little or no solid solution with B. Then on the phase diagram, A defines virtually a 'line phase', flanked to either side across two-component spans by inhomogenous two-phase behaviour. Iron monoselenide would appear to find itself in such a circumstance in relation to potential non-stoichiometries $Fe_{1-\delta}Se$ and $FeSe_{1-\epsilon}$. $Fe_7Se_8$ is a well-established phase along the former path. The unfortunate thing about our target composition, FeSe, is that this itself actually falls within a two-phase range of composition. A line phase (or virtually so) does in fact separate the two non-stoichiometric and indeed two-phase ranges linked above with 'iron selenide', but the stable homogeneous single-phase material emerges as being slightly selenium-poor at $FeSe_{0.974}$ [108 fig.5] (or equivalently, so far as the mass ratio goes, iron-rich at $Fe_{1.02}Se$ [53a fig.3b]). Stable non-stoichiometric 'defect compounds' of this type are not uncommon among transition metal materials, the superconductor 'TiO' being an extreme case in point. Even GaAs is, note, very slightly off-stoichiometry as crystallized from its melt. Phase separation is not confined just to the iron selenide superconductor [53a,108], but appears now widespread amongst the pnictide systems too [114;(Ba/K)$Fe_2As_2$],[115;Sm(O/F)FeAs]. Between pure and substituted conditions clearly quite strongly repulsive interactions must arise to introduce 'dimpling' so ubiquitously into the free energy curves. Such materials problems can be a real menace when trying to deal with an already complex and delicate situation, but maybe with regard to the electronic problem now in hand one can turn this complexity of form to some advantage.

There arises comparably intricate and widespread action in regard to the orthorhombic distortion and its precursor thermal behaviour. Again the thermodynamics can suggest processes and provide illumination as to what is going on. We cited earlier the case of the thermal expansivity data from Klingeler *et al* [107] for members of the 1111 family. The effect of the T→O transition (at $T_D$) upon $\alpha(T)$ was to cause a positively signed anomaly, whilst, conversely, the effect of the SDW onset (at $T_M$) upon $\alpha(T)$ is to generate a negative anomaly. Application of the Ehrenfest thermodynamic relation necessitates then that $dT_M/dP$ will see reduction in $T_M$, whereas pressure increase will bring a rise to $T_D$, thereby increasing differentiation between the two transition temperatures. Any potential *tetragonal* CDW will have lost out here in stability range to the 1D orthorhombic state, while the ordered magnetic state likewise will have become more confined in phase space.

In looking at the detailed outworking of these matters let us first begin with FeSe, before delving into the extensive substitutional changes of the 1111 and 122 families.

In FeSe the Fe-Fe distance is only 2.66 Å, as against 2.80 Å in $BaFe_2As_2$ and 2.85 Å in LaOFeAs. At 90 K there takes place in 'FeSe' a loss of structural symmetry much as occurs in $BaFe_2As_2$ at 138 K and at 150 K in LaOFeAs; namely a slight distortion of the original tetragonal structure into an orthorhombic, √2x√2 supercell, wherein $b/a$ duly comes to differ from unity by just ~½%, angle $\phi$ (see figure 9) departing from 90° by only around 0.3°. Recall from what was said though in §5, despite such small modifications to the cell itself, the atomic shifts internal to the unit cell need not be insignificant. Indeed in $Fe_{1.01}Se$ a splitting in Fe-Fe separations of 0.012 Å is acquired by 20 K [116], while in LaOFeAs the difference is marginally greater at 0.014 Å [117]. The really striking distinction between the two materials, however, is that with FeSe no magnetic transition follows, unlike with stoichiometric LaOFeAs, etc. for which elastic neutron scattering work finds a magnetic moment ~$0.3\mu_B$ ordered in the spin arrangement of figure 9. No such SDW development is experienced with near-stoichiometric $Fe_{1.01}Se$, the Mössbauer trace reported in [116] as elsewhere being throughout of a magnetic singlet state. The dominant RVB-type spin-pairing implicit here in the selenide has to be a reflection of its much closer Fe-Fe spacing. LiFeAs, for which Fe-Fe is 2.68 Å very much as with FeSe, indeed displays no SDW in advance of $T_c$ (= 20 K) [6,118], and, what is more, seemingly no orthorhombic distortion either, though that now needs closer examination by TEM. (Note in LiFeAs $\alpha$ is only 102.9° [6b]).

Since $Fe_{1.01}Se$ presents its orthorhombic distortion without any SDW the distortion clearly cannot be magnetostrictive in origin. A CDW automatically is indicated. Let us examine more closely then the detailed form of the PLD suggested in [116] from their combination of low temperature synchrotron X-ray diffraction and transmission electron microscopy. The latter experiments reveal that the distortion in fact is of lower symmetry than the customarily quoted Cmma form (space group #67, $D_{2h}^{21}$). The latter group simply would express the √2 expansion from the parent tetragonal cell (space group #129, P4/nmm) plus the newly acquired cellular orthorhombicity. However the diffraction results and in particular the TEM data make it here apparent, through the occurrence of 110-type super-spotting, that the correct space group has to have shed the horizontal 45° glide plane 'a' (the former 'n'-plane of the tetragonal parent). The observed absence of 100-type super-spots would, however, signal the retention of C-face-centring. That there no longer exists the half-integer horizontal glide, linking neighbouring Fe atoms in the $x_O$ direction in the basal plane supports the fact that those atoms have become slightly dimerized as intimated in the previous paragraph. The situation expressed by the data is as conveyed above in figure 11c. The space group looks to have contracted here to #35, Cmm2 $C_{2v}^{11}$, although the actual loss of symmetry could in reality be somewhat more severe as the selenium atoms accommodate to maintain Fe-Se bond lengths. Note the arrangement of fig. 11c betokens a bond-centred CDW as against some site-centred restructuring of the charge distribution. It is a prime expression of RVB coupling having tied up the residual spin into now static, non-magnetic,

S=0 dimer entities – a scaled-back version of what is encountered in $d^5$ arsenopyrite FeAsSe (see §2). Of course for our material in due course to exhibit superconductivity necessitates that such 'preformed' pairs are or can be rendered electronically mobile, and that a semiconducting gap of the severity developed in FeAsSe, $Ti_2O_3$, etc. is avoided.

The above may well provide the key to understanding the additional striking discovery made in [116]. $Fe_{1.03}Se$, placed at the opposite edge of the 'line phase' to $Fe_{1.01}Se$, proves remarkably different from the latter in its properties. Not only does it not become orthorhombic at low temperatures but, furthermore, not superconducting either [116]. However what McQueen *et al* do observe from their TEM work on $Fe_{1.03}Se$ is that its structure clearly is not covered by tetragonal space group P4/nmm, any more than it is either by orthorhombic groups Cmma or Cmm2. It would appear from the simultaneous presence of both 100- and 110-type superlattice spotting that a strictly √2x√2 enlarged tetragonal cell has been taken up. Direct lattice imaging of the sample reveals intensity fringing with a basal spacing of 5.4 Å, or twice what is witnessed with $Fe_{1.01}Se$. It is just feasible that what is being registered here for the current $d^6$ electron-count system is an array of spin quadrimers, as introduced in §8 and portrayed in figure 11a. Strong, high-order, structural groupings are a feature of many transition metal compounds under the appropriate electron count; e.g. the quadrilaterals in $d^3$ $ReS_2$ [1] and the triangles in $d^2$ $LiVO_2$ [119], with weaker such groupings, as noted earlier, occurring in hexagonal FeS [12] and NiAs [120]. The TEM diffraction plate included by McQueen *et al* [116 fig3d] shows, from the appearance of both 100-type and 110-type spotting, that the basal glide and C-face-centring elements have each been abandoned in the prevailing space group, although the latter clearly retains overall tetragonality. Group $^{bar}42m$, #111 $D_{2d}^{1}$ would accommodate these changes, the Fe atoms being located here in its four-fold degenerate, type m sites. Perhaps this quadrimer condition is now too correlated, too non-metallic, for the semimetal to uphold superconductivity (at least above the 0.5 K checked to). Certainly the resistivity is significantly higher here than even with $Fe_{1.01}Se$ [53a fig.5].

Interestingly a comparable distinction in diffraction and resistivity behaviour has been picked up by Wang *et al* [121] using thin films of 'FeSe' grown epitaxially on MgO (provided the films are unrelaxed, possessing thicknesses below 150 nm). When grown with the substrate held at 500°C the product is just as for bulk $Fe_{1.01}Se$, but if the substrate is held at only 320°C the behaviour of the film becomes much more like $Fe_{1.03}Se$, with no low temperature orthorhombicity and no superconductivity. Strikingly a growth habit difference occurs between the two products. The 320°C product, from the $a_o$ fringing present in its TEM lattice image, involves a somewhat strained (expanded), non-face-centred, tetragonal [001] growth mode, whereas the 500°C product shows [101] growth with fringes perpendicular to [$1^{bar}11$]. As stated, the latter product exhibits both orthorhombic distortion and superconductivity at low temperature; the former does neither. It is crucial now to discover how TEM imaging and diffraction from specimens like these alters upon cooling below room temperature.

A similar data gap exists for the temperature evolution of the Fe site Mössbauer signal forthcoming from the two types of product. Precisely how does the quadrupole splitting in these materials alter with $T$? Figure 6 in [53a] from McQueen *et al* would suggest a distinction between the two in the quadrupole-split line-broadening observed at 5 K as compared with 295 K, it being greater in $Fe_{1.03}Se$ than in $Fe_{1.01}Se$. The Seebeck data in the same paper [53a inset to fig.5] seem to signal the possibility of an event at 130 K in the case of $Fe_{1.03}Se$. There the magnitude of $S$ becomes largest, at $-16$ μV/K, this following (for both compositions) a sign change from + to – below 230 K. The results in [53c fig. 4] are very similar. Is there any further indication in the literature of some 130 K event in less closely characterized 'FeSe'? Figure 7 in [122] would suggest the occurrence of a slight alteration there to the change in $c/a$ with $T$, whilst figure 3a in [123] claims to see a sharp discontinuity at this temperature in paramagnetic magnetization.

At this point the reader perhaps feels the distinction between $Fe_{1.01}Se$ and $Fe_{1.03}Se$ is impenetrable - or worse - not worth penetrating, a detail ascribable to 'dirt'. However one very powerful microprobe of any complex situation is always NMR. In reference [24d] from Imai *et al* we already have observed how in FeSe the Knight shift signal from $S = ½$ $^{77}Se$ suffers a strong reduction in magnitude with temperature, beginning right from the phase stability limit of 500 K. This reduction in $K_s$ (or equivalently $\chi_{spin}$) is virtually identical for both $Fe_{1.01}Se$ and $Fe_{1.03}Se$ [24d fig.3]. There is a marked difference between the two, however, when it comes to measuring the *integrated* intensity of the NMR line. With $Fe_{1.03}Se$ the latter is invariant below 100 K (although the line broadens somewhat), but with $Fe_{1.01}Se$ a clear drop off in *intensity* is recorded coming *in advance of* $T_c$. This distinction becomes much more marked under applied pressures of up to 2.2 Gpa (for which $T_c$ does not exceed 15 K ) – and it occurs without change to $K_s$ itself. How can 50% of the NMR signal have disappeared in $Fe_{1.01}Se$ under these conditions, yet not in $Fe_{1.03}Se$?

That this 'wipeout' is fluctuational in nature immediately is made evident upon turning to the relaxation rate data. $Fe_{1.01}Se$ below ~130 K deviates from its hitherto linear-in-T fall in relaxation rate $1/T_1$, and $1/T_1T$ actually turns up sharply below 90 K ($T_D$) to produce a marked hump. The latter hump (when integrated) becomes more pronounced under pressure and always occurs at a T in advance of the growing $T_c(P)$ [24d fig 4]. The fluctuations have widely been viewed as due to spin fluctuations. As these fluctuations take on RVB form and freeze out as dimer pairs the susceptibility experiences 'wipe out'; the system becomes spin pre-paired and ready to move forward to the special superconductivity of these materials [24d fig.3]. $Fe_{1.03}Se$ lacks all aspects of this fluctuation freezing. With $Fe_{1.01}Se$ under pressure as the freeze-out arrives at higher $T$ so too $T_c(P)$ moves up. Note this spin freezing is not spin ordering because that would bring a splitting of the Knight shift spectrum, which is not seen.

A related technique to examine the local magnetic condition on short time scales is μSR. Being an expression of precisely where the muon localizes prior to decay, the technique enables one to assess the homogeneity of the sample via observation of the spin depolarization rates for the muons as they precess, either in a small applied magnetic field, or in a self-generated field. Below $T_c$ and for fields greater than $H_{c1}$ the sample is penetrated by flux vortices wherein the field, being non-uniform, leads to non-synchronization of the individual site precessions, and hence decay of the overall precession signal. This depolarization rate is directly related to the penetration depth, and hence to the local superfluid density $n_s$. In a significantly phase-separated situation one is in a position then to pick up multiple signal decay rates.

Such phase separation quickly was registered by following µSR decay in a field transverse to the spontaneous field direction when using a sample of slightly 'underdoped' (Ba/K)Fe$_2$As$_2$ ($T_c$ = 37 K) [114a]. There it was deduced that only 25% of the sample actually was superconducting, with 50% locked into the SDW condition, proportions that change with $T$. To confirm such a phase separation, zero-field scanning magnetic force microscopy was undertaken (at 26 K), and the granularity of the condition became directly evident. By Fourier analyzing the scanned image it was found that the typical span of the magnetic segments was about 650 Å, which is not too dissimilar from $\lambda$ (and much greater than the stripe domaining of the cuprates). Both powder and single crystal neutron diffraction revealed this *electronic* phase separation to be occurring in material which crystallographically appears homogeneous and orthorhombic. There is comparable evidence the filamentary superconductivity recorded in Na$_{1-x}$FeAs occurs under very similar conditions [124].

Besides examination of two-phase structuring, µSR also allows equally valuable information to be extracted from the present materials below $T_c$ in relation to the progressive development of the superconductivity across the semimetallic divide. Under pressure it proves possible specifically to track the developments of the penetration depths and the superconducting gaps as functions of $T_c(P)$. Using the theory introduced by Kogan *et al* [125] for a two-gap superconductor, Khasanov *et al* have been able to resolve the strikingly different behaviours exhibited in the Γ pocket and the zone corner pocket. Compressed powder samples of nominal iron selenide compositions FeSe$_{0.94}$ and FeSe$_{0.98}$ were measured and analysed by Khasanov *et al* [126] in this two-pocket manner. Over the pressure range up to 0.85 GPa, $T_c$ rises smoothly from 8.5 K to 13 K. Here $T_c(P) \propto \lambda^{-2}(0,P)$, as foreseen by Schoenberg. When decomposed into its two components, the smaller (X point) gap hardly alters in its $\Delta$ or $\lambda$ (i.e. $n_s$) contributions to the sum, but for the larger (Γ point) gap these quantities each augment linearly as $T_c(P)$. The two components behave almost independently, the contribution of the smaller gap arriving only well below $T_c$, but the two sets of carriers still are sufficiently coupled that just the one overall $T_c$ applies [126 fig.2]. This is a situation very like the relationship that the plane and chain carriers possess in YBa$_2$Cu$_3$O$_7$, where the additional superconductivity of the chains boosts $\lambda^{-1}$ and $n_s$, but brings virtually no boost to $T_c$ for the entire ensemble. Analogous behaviour recently has been reported for optimally-substituted Sm(O/F).FeAs by Weyeneth *et al* [127] employing a combination of torque magnetometry, SQUID magnetometry and µSR. As the applied field is increased to ~1 T (a field way below $H_{c2}$ for the material *in toto*) it is observed that the larger gap, as expected, is virtually field-independent, but the smaller gap is strongly suppressed. These experiments demonstrate conclusively that in these materials the high $T_c$ behaviour is emanating from the RVB pairings of the zone centre pocket.

The above, in respect to FeSe, related to $P$ being less than 0.8 GPa. Above this pressure a whole series of further new and highly illuminating phenomena arise. It for some time has been known that $T_c(P)$ shows a hiatus in its growth, with an actual minimum showing up in $T_c(P)$ between 1 and 2 GPa [128], prior to renewed rise to attain a $T_c$ which peaks at over 35 K in the best samples under a pressure ~10 GPa [26b]. The latter high pressure range is free from any pair-breaking scattering due to magnetic order, as is apparent from the Mössbauer spectrum there [26b fig.2b]. The latter is of singlet form. Beyond 6 GPa there is, though, clear evidence of two-phase behaviour, with the development of a component that appears non-metallic. What, however, of the pressure range between 1 and 2 GPa, within which the $T_c$ minimum occurs, and for which no Mössbauer results are available?

Upon returning to the μSR results of [126] for this range one discovers that it is a region in which a very large fraction of the overall sample opts to develop magnetic order. This order evidently coexists intimately with the superconductivity because the magnetic order parameter (onsetting between 20-40 K) drops away appreciably in amplitude as the superconductivity arrives near 12 K [126 fig.3]. Why should pressure have this interim effect of introducing magnetism to FeSe in this way? The answer has to lie in the fact that FeSe is a true layer compound, where $c_o$ initially drops disproportionately rapidly with $P$, falling from 5.5 Å to 5.1 Å over the above pressure range [26b fig.1d]. This decrease promotes the 3D magnetic coupling strength greatly. Ultimately the slower reduction in $a_o$ and the ensuing increase in the direct basal Fe-Fe interaction tips the balance back in favour of RVB, and the superconducting $T_c$ can rise again. It should be recalled here that FeSe has the smallest $a_o$ of all these materials, even prior to the application of pressure, and now by 10 GPa it has declined a further 4%. Where such interaction between the competing superconducting and magnetic order parameters has similarly been directly recorded is in the small coexistence range existing in Ba(Fe$_{1-x}$Co$_x$)$_2$As$_2$ near $x$ = 0.05. The latter observation was made by neutron diffraction, measuring specifically the effect on the magnetic order parameter. As Fernandes, Pratt and coworkers [129] demonstrate in a full analysis of their data, it is observations such as this which confirm the superconducting order parameter to be here s$^{+-}$ and not s$^{++}$.

**§10 Superconductivity of note in other homopolar bonded, semimetallic materials.**

The features which turn FeSe, etc. into remarkable superconductors might then be listed as follows -

(1) homopolar bonding that brings considerable movement of the associated bonding and antibonding bands; (2) a moderately complex crystal structure for which the symmetry is not too high and there result quite a number of bands; (3) a low-dimensional structure, here layered (but perhaps a chain or even an individual cluster structure would suffice), that affords ready intersite spin dimerization; (4) a moderately ionic system where strong local characteristics are still preserved and the carriers are not too delocalized; (5) one, however, in which magnetism does not dominate; (6) a semimetallic situation to which structural dimerization has not brought the complete removal of free carriers; (7) a semimetal in which the overlap is indirect, bringing into play the instabilities of the excitonic insulator and boson-fermion degenerate tunnelling; (8) a semimetallic overlap that is not too great, limiting the level of screening of local action and also restricting Fermi surface nesting.

How has FeSe acquired such characteristics?

Feature (1) comes from a structure in which the tetrahedral Fe coordination units share edges, bringing quite a short Fe-Fe nearest-neighbour separation.

Feature (2) comes as the adopted structure is a non-symmorphic, tetragonal one in which $Z$=2 (unlike its zincblende polymorph).

Feature (3) represents a rather unusual circumstance for a monochalcogenide of a transition metal element, of being marginally stable here over the nickel arsenide structure, this seemingly on account of the current $d^6$ electron count and what this engenders near $E_F$.

Feature (4). FeSe and even FeAs might still be classified to some degree as retaining ionic characteristics, evident from comparison with GaSe or AlAs.

Feature (5). The NiAs-structured, S=2 polymorph of FeS is strongly magnetic and metallic, characteristics avoided in our current system.

Feature (6). Feature (5) came with the semimetallic, tetrahedrally bonded product. With the number of delocalized carriers rather low, the drive to dimerization is not then overwhelming.

Feature (7) is the product of the basal glide-plane in the adopted structure. The bosonic carriers degenerate with the C.B. fermions lie at the top of the M-M bonded V.B. at the $\Gamma$ point, and are of low crystal momentum. The valence bands in question show virtually no dispersion along $\Gamma Z$, and at coordination unit shape ideality become locally spin-isotropic. It is for the bosonic pairs from these zone centre bands that superconductive gapping is high.

Feature (8). If the semimetallic overlap becomes too large as in FeTe, or too small or even absent as in FeS, then the superconductivity becomes less striking or vanishes altogether.

Quite a lot of materials share of course many of the above attributes, and some indeed are superconductors that have caught attention in the past. As noted earlier, the A15's like $V_3Si$ and $Nb_3Ge$ are semimetals which display martensitic dimerization, but this arises in a 3D cubic setting and the semimetallic overlap at $\Gamma$ is direct [130]: $T_c$ is limited to 23 K. The Chevrel phase superconductors like $PbMo_6S_8$ looked promising too. The $Mo_6$ M-M bonded octahedral cluster that supports $E_F$ is however built into a 3D (cubic) structure, and the units are quite widely spaced. The $A_3C_{60}$ Bucky Ball superconductors are in a not dissimilar situation, with alkali atoms intercalated between the $C_{60}$ clusters to produce an open band situation. As was expressed in [131], I believe that the right way to address superconductivity in $A_3C_{60}$ is of *pair* tunnelling between the cluster balls. The $C_{60}$ units, composed of weakly distinguished double and single bonds, encourage negative-$U$ pre-pair formation, and the alkali ions set up a significant local perturbation on the structure and electronic environment. I expressed it a big mistake at the time to proceed with a rather standard MacMillan and Dynes type treatment as pursued by Schluter *et al* [132] and others.

Graphite is a semimetallic polymorph of carbon with a very limited carrier population. The band overlap here is direct, between $p_z$-$p_z$ $\pi$-bonding and -antibonding states, and superconductivity when derived from this base by intercalation is only of low $T_c$. Where things have become much more exciting has been with $MgB_2$, a pseudographite intercalated by $Mg^{2+}$ ions. Here both $\sigma$- and $\pi$-bonding states are involved in a situation in which the semimetallic gap now is indirect. A $T_c$ of 40 K still has not deterred people from proceeding with a standard treatment in $MgB_2$, despite it being clear specific phonons contribute to coupling constant $\lambda$ in a very anharmonic fashion [133]. It would seem $MgB_2$ is coming very close to kinship with FeSe. In fact all the above materials actually find accommodation on or close to the Uemura plot, wherein high $T_c$ is acquired for remarkably small $n_s$ values, as determined from penetration depth/$\mu$SR studies. Where the iron pnictide materials would seem to surpass $MgB_2$ in performance is that the homopolar bonded states in play at $E_F$ are transition metal states and involve, beyond the primary $\sigma$-interaction, secondary $\pi$-interaction within the iron atom sublattice. This permits now a more variable density of states and somewhat greater flexibility of action, provided, that is, one can avoid the pair-breaking action of ordered magnetism. It could be thought one way to achieve the latter would be to turn to 4d and 5d materials, but this could lead to dimerization becoming too strong (witness $NbO_2$ vs $VO_2$), or to spin-orbit splitting introducing unwanted instabilities. It is notable, however, that the current iron pnictides will tolerate a considerable replacement of Fe by Ru without $T_c$ falling too strongly, despite the broadening of the bands [134].

In truth the window of opportunity afforded by SmO.FeAs, etc. is a very narrow one, and although we now see a plethora of new 'high $T_c$' superconductors, as openned up in [95]-[99] with $(Fe_2As_2).(Sr_4(Mg,Ti)_2O_6)$, etc., these all prove variants on a very restricted theme. As was seen from figure 1 regarding the 1111-family, the response here to slight change is supersensitive and fascinating to follow. However the situation would appear not to offer great hope for emulating the remarkable heights reached in the equally sensitive cuprates with the resonance on show in $HgBa_2Ca_2Cu_3O_{16+\delta}$, and $T_c$ under pressure of 165 K.

In closing three more contact points should be made. Firstly the excitonic insulator condition alone does not suffice to yield high $T_c$ on the evidence of what is observed in layered $TiSe_2$ [135][136] or pyrite $\beta$-$SiP_2$ [137]. Secondly $\beta$-ZrNCl (which is quite differently layered from the $\alpha$-form), containing a strongly Zr-Zr bonded Zr bilayer and supporting superconductivity around 25 K when intercalated with Li [138][139], witnesses its superconductivity to maximize just prior to Mott localization [140]. (The $\alpha$-form is a simple salt). Notably this 4d-series superconductor again falls on the Uemura plot [55].

Finally we have the question of where we are proceeding with the remarkable superconductivity manifest in many elements under high pressure. For some time now we have been made aware of raised $T_c$ superconductivity in group-B elements like silicon (12 K) [141], phosphorus, or sulphur in their high pressure polymorphs. Under pressure these elements move away from their low pressure often molecularized non-metallic structures to acquire low-dimensional network structures, bi-layered in the case of black phosphorus ($T_c$=14 K). With sulphur $T_c$ reaches a value of 17 K [142]. All these superconducting forms are semimetals, and of course are homopolar bonded, retaining still a small nearest-neighbour count rather than shifting to simple 'close-packing'. ('Close-packing' is, note, 12 n.n atoms standing off at considerable uniform separation.) The inverse of the above process we now find happening in group I, II and III metals. These elements transfer under pressure into reduced coordination number structures – the sort of change existing between Al and Ga. The resulting structures can become very complex, sometimes incommensurate, or of the 'guest and host' type, a bit like boron [143]. Typically the phases where superconducting are semimetallic in form. We witness Li acquire a $T_c$ of 20K [144], and Ca and Y reach a $T_c$ up around 25 K - and still rising under pressure [145]. Once more it appears we are seeing aspects here of what we have been discussing earlier in regard to the iron pnictides.

The message definitely in all this is to examine more low-dimensional, homopolar-bonded semimetals - say perhaps $CaSi_2$ or a gallium or titanium boride cluster system.

**§11 Summary**

The pnictide superconductors are indirect gap semimetals, appreciably more delocalized than the cuprates. Their unusual form of layered, tetragonal crystal structure is responsible for the semimetallic outcome. The latter is secured largely in consequence of the strong direct Fe-Fe bonding/antibonding interaction experienced here. This direct interaction leads to materials wherein the magnetic spin becomes locked, not in ordered magnetism, as with nickel arsenide structured FeS or FeP, but in RVB fashion. On cooling, the paramagnetic susceptibility becomes quenched linearly with falling temperature, and when/if ordered magnetism becomes realized at reduced temperatures it takes a low effective moment. Such ordered magnetism (plus its spin excitations, so damaging to superconductivity) can be suppressed by adding carriers to the materials, or often by applying pressure, increasing quasiparticle delocalization.

Being indirect gap semimetals, the materials are susceptible to excitonic insulator type electronic instabilities adopting the wavevector separating the electron and hole pockets. The magnetic order actually to emerge takes up this specific wavevector. Whenever such magnetic order appears, it always is prefigured by a slight structural symmetry breakage to orthorhombicity, this likewise displaying the above instability wavevector. Neither magnetic order nor orthorhombic distortion prove necessary forerunners to the superconductivity, LiFeAs being a case in point. All that is required is for a suppressed magnetic susceptibility, imposed by the dominant RVB coupling.

The dynamic bonding spin-pairings of the carrier states around $\Gamma$ are seen as constituting bosons, these resonant with the decoupled fermions in the quasiparticle pocket at the zone corner. The ensuing boson-fermion transfer is viewed as being responsible for the elevated $T_c$ superconductivity of these systems. The superconductive gapping acquired is significantly different on the $\Gamma$ and X point pockets, it being considerably greater on the former, from where the population of bosonic pairings originates and the action is governed.

The particular materials able to sustain the highest $T_c$'s are those for which the paramagnetic susceptibility is small and the site moment most constrained, and for which the coordination geometry is moving toward being perfectly tetrahedral. The latter shape ideality has the effect of procuring degeneracy between the $d_{xy}$, $d_{yz}$ and $d_{xz}$ Fe-Fe pairing states, strengthening RVB, advancing correlation, openning a pseudo-gap, and promoting the excitonic insulator instability in a manner such as to favour superconductivity over an SDW or CDW end product.

The above three are competitor ground states. They can occur in conjunction, but then impair each other's order parameters, competing for the same electrons over a Fermi surface very limited in extent. $T_c$ currently is greatest in fluorine-substituted SmO.FeAs, a system where shape ideality is close to being achieved, but which has to accept a considerable $e/h$ imbalance in order to inhibit magnetic ordering. If it were possible to accomplish the latter via intercalation by a non-donor such as $H_2O$ it might be possible to see $T_c$ rise somewhat.

Superconductivity is not as spectacular with the pnictides as with the cuprates because the bosons generated by RVB coupling are not so robust as the carrier pairings created under the negative-$U$ double-loadings within the mixed-valent environment of the HTSC cuprates. The pnictides, nevertheless, hold a good many features in common with the cuprates which can make the systems appear more similar than they truly are. In each case the strong electron-boson scattering renders the conductivity near-incoherent at raised temperature. The presence of a well-defined bosonic state leads to comparable IR optical spectra, and their spin excitation under spin-flip inelastic neutron scattering once more leads to a characteristic 'resonance peak'. Additionally there is the very similar NMR spin-lattice relaxation behaviour, particularly below $T_c$, while, of course, both families cluster about the Uemura plot, as revealed in μSR penetration depth studies.

With both families, since the controlling interaction is electronic, $T_c$ is easy to manipulate by substitution or intercalation, and either set of materials readily descends into complicated micro-structural behaviour, although in the pnictides there is nothing quite like stripe phase formation. For the cuprates such structuring has a strong coulombic input, but the pnictides are more covalent and better screened dielectrically, despite their low carrier count. In the end the more ionic character for the cuprates stands as what upholds their remarkable, unique, negative-$U$ state behaviour, and from this their truly high $T_c$ superconductivity.

**Acknowledgement.** The paper is dedicated on the occasion of his ninetieth birthday to Dr Abe Yoffe of the Cavendish Laboratory, Cambridge - my former Research Supervisor and long term Editor to the forerunner of this Magazine.

**Pnictide Ref Index**

**Figure Captions.**

**Figure 1.** (p5). A plot of bond splay angle, $\alpha$, within the bisphenoidal Fe coordination unit versus achievable $T_c$, for a wide range of iron arsenides and phosphides. The plot shows the the maximum value $T_c^{max}$ attainable superconductors (at 1 atmos.) with each system included, and it features the sharp global maximum of 56 K realized within the SmOFeAs system as $\alpha$ transits the value of 109°28' for a regular tetrahedron. Note the progression across this peak traced out by the rare-earth series in 1111-Ln arsenides, corresponding to growng basally directed chemical pressure upon the FeX sandwich from right (La) to left (Dy). The plot is taken from Lee *et al* [27] [reproduced here with permission of *J. Phys. Soc. Jpn.*: copyright 2008]. Note additionally the more recent results -

| 1111-Ho [27b]; -Er | SrFe$_2$As$_2$ [8a, 146] | LiFeAs [6] | FeSe (1 Atm.) [26b] |
|---|---|---|---|
| $T_c$ = 36 K; 33K. | $\alpha$ = xxx°, $T_c$ = 6–26 K. | $\alpha$ = 103°, $T_c$ = 21 K. | $\alpha$ = 105°, $T_c$ = 11K. |

[27b] Rodgers J A *et al*, 2009 *Phys. Rev.* B **80** 052508.  [146] Kim J S *et al*, 2009 *J. Phys.:CM* **21** 342201.

**Figure 2.** (p5).
**(a)** Bisphenoidal orientation of tetrahedron present in mackinowite FeS. There occurs here slight elongation along *z* as compared with a regular tetrahedron. The Fe atom resides halfway between the upper and lower crossed edges of the coordination unit. The latter run parallel to the *x,y* axes of the tetragonal cell. The units share all four remaining edges to build a layer structure.

**(b)** Basal projection of the bisphenoid unit. The inner, dotted square marks the horizontal section through the central Fe site, and the *X*,*Y* axes shown form the natural symmetry axes of the unit. Directed to the mid-points of the inclined tetrahedral edges (and beyond in the full structure to the n.n. Fe sites) they run at 45° to the crystallographic axes *x,y* for the mackinowite structure.

**(c)** Side elevation of the layered mackinowite form of FeS, in which the FeS sandwiches sit directly above each other across a somewhat contracted vdW gap. Bond-splay angle $\alpha$ monitors the level of deformation of the tetrahedra from being regular (109° 28').
N.B. $\sin(\alpha/2) = \frac{1}{2}a_o/B$ and $\tan(\alpha/2) = \frac{1}{2}a_o/U$, where $B$ is the FeS bond length and $U = u_z(S).c_o$. (In FeS $a_o$ = 3.678 Å, $c_o$ = 5.039 Å, and for $u_z$ = 0.258, $\alpha$ would be ideal. In FeSe $u_z$ = 0.268 making $\alpha$ = 104° [24a].)

**(d)** Single FeS sandwich of mackinowite structure in plan. The form of the sandwich is like a waffle with square pyramidal pits that just penetrate its thickness. The pits on the reverse side are staggered by (½,½). The *Z*=2 crystallographic unit cell, tetragonal space group $D_{4h}^7$, P/nmm, #129, requires that in 'first setting' the Fe atoms lie at the cell origin ($^{bar}$4m2). The nearest-neighbour Fe distance between the atom at the corner of the f.c. Fe planar array and that at its centre is just 2.60 Å. The alternative 'second setting' sees a transfer of the cell origin to the centre of inversion at ($^{bar}$¼,¼).

**Figure 3.** (p7). Molecular orbital level scheme for mackinowite FeS, covering the S 3*p*- and Fe 3*d*-derived states, this directly bearing on the band structure at the $\Gamma$ point. The figure indicates how the levels for an isolated tetrahedral coordination unit, sphenoidally oriented in the crystal field, become first modified from the customary levels of the zincblende form and then further much modified in the current *Z*=2 structure under the very strong direct Fe-Fe interaction (separation just 2.60 Å). The parentage of the states is indicated in the centre of the figure, whilst the symbols to the right give the representations appropriate to space group $D_{4h}^7$, to appear on the band structure below. The numbers in brackets mark the various state electron capacities. With the present complement of 12 *d* electrons for the *Z*=2 Fe(II) condition, the closely spaced pair of states $3^+$ and $5^+$ which issue from $d_{xy}$ and $d_{xz}, d_{yz}$ respectively will constitute the top of the V.B.. Note both the latter are $pd\sigma^*$ states driven downwards below $E_F$ by the Fe-Fe bonding. This addition to the V.B. is partially countered by analogous strong elevation above $E_F$ of the $dd\pi^*$ antibonding partner issuing from $d_{x2-y2}$, so fixing the energy gap for $d^6$ FeS to be between the $\pi$ and $\pi*$ states which emerge from $d_{xy}$. Note above we are using state designations relative to the crystallographic axes *x,y*, not the coordination unit axes *X,Y*. Figure based on that of Welz and Rosenberg [33]: [copyright: IOP, 1987]. The energy scale has been transcribed from Rydbergs to eV and the zero level shifted very slightly to coincide with the top of the valence band.

**Figure 4.** (p9). LMTO band structure for tetragonal FeS obtained in the A(tomic) S(phere) A(pproximation) and using the L(ocal) D(ensity) A(pproximation to treat correlation and exchange. The figure comes from the paper by Subedi *et al* [31]; [copyright: APS 2008]. Shading has been added to indicate the extent of direct gapping between the *p*- and *d*-bands, and similarly with regard to the intra-*d*-band gapping about $E_F$ under the *Z*=2 cell's 12-electron complement at Fe(II) $d^6$ loading. The figure has been amplified to show the symmetry representations at the special points in the zone, this information extracted from the paper by Welz and Rosenberg [33]. Note the compulsory band degeneracy arising throughout the cuts XM, MA and AR due to the presence of the horizontal glide plane.

**Figure 5.** (p9). Band structure for LaOFeAs obtained by Vildosola *et al* [35] using LAPW method and following the WIEN2k routine within LDA. The band structure is strikingly similar to that of figure 4 for tetragonal FeS. The shading about $E_F$ makes evident the small indirect overlap between the VB maximum at $\Gamma$ and CB minimum at M, the cell corner. The widths of the occupied and unoccupied sections of the d-band are here somewhat over-stated, making this overlap too large. The calculation enfolds the *experimentally* derived value for $\alpha$ of 114.6°. [Reproduced with permission of APS: copyright 2008]

**Figure 6.** (p9). Key result from Kimber *et al* [36] for the $BaFe_2As_2$ system showing the identical behaviour of the structure under hydrostatic pressure and under K-doping. Both agencies secure, following elimination of the orthohombic distortion, a growth and then

decline of $T_c$ over a maximum reached in the pressure sweep at 4 GPa and in the doping sweep at 35% K. The plot reveals that under *both* procedures the maximum $T_c$ value materializes precisely as the tetrahedron passes through the *regularized* inter-bond angle, 109.44°. [Reproduced with permission of MacMillan press: copyright 2009].

**Figure 7.** (p9). A blow-up of figure 5 for LaOFeAs ($\alpha$ = 114.6°) in the vicinity of $E_F$, for comparison with the corresponding result for LaOFeP ($\alpha$ = 120.2°). The semimetallic overlap is smaller in the arsenide and the $d_{xy}$ and $d_{xz},d_{yz}$ states in the $\Gamma$ pocket have come closer to triple degeneracy as $\alpha$ is brought toward ideality. [From Vildosola *et al* [35]; reproduced with permission of APS]

**Figure 8.** (p11). Side elevations, approximately to scale, of the tetragonal FeSe, LiFeAs, LaOFeAs (all in 'first setting') and BaFe$_2$As$_2$ $Z$=2 unit cells.

| LaOFeAs (1111) | LiFeAs (111) | BaFe$_2$As$_2$ (122) | FeSe (11) |
|---|---|---|---|
| P4/nmm | P4/nmm | I4/mmm | P4/nmm |
| $a$ = 4.03 Å | $a$ = 3.77 Å | $a$ = 3.96 Å | $a$ = 3.77 Å |
| $c$ = 8.74 Å | $c$ = 6.36 Å | $c$ = 13.02 Å | $c$ = 5.52 Å. |

**Figure 9.** (p17). Illustration of the magnetic cell and the spin order assessed by neutron diffraction to hold for some way below $T_d$ in many of these materials. The √2-by-√2 orthorhombic magnetic cell is indicated centred about a Fe site in a parent tetragonal cell (presented here in 'second setting' to emphasize the equivalence of its two Fe atoms). In the portrayed antiferromagnetic array such order would be in keeping in a local modelling with superexchange coupling interaction $J_2$ borne through the intervening As atoms stationed in the sheets above or below the Fe sheet as the case may be. In the low temperature condition the dominant interaction becomes direct intracellular spin pairing between the two Fe atoms there. Near $T_c^{max}$ in the spin-gapped conditions so developed no magnetic LRO ultimately remains, and an RVB state is acquired in which all spin alignment orientations have become equivalent. The observed orthorhombicity recorded with almost all of these materials at low temperatures is thus necessarily not the consequence of magnetostriction – as indeed it was not above $T_{SDW}$ either. (Following convention the axes for the orthorhombic cell are labelled such that $b_O$>$a_O$.) Throughout as $T$ goes down the static susceptibility declines linearly.

**Figure 10.** (p24). A view of the state of play in the $T$ vs $x_{Co}$ field for the system Ba(Fe$_{1-x}$Co$_x$)$_2$As$_2$ secured by Ning *et al* from $^{75}$As NMR work [90]. The two heavy lines mark the onset temperatures of the SDW and of the superconductivity. The former state is lost as the latter is approached. At low cobalt content the superconductivity looks as if it would reach a $T_c$ ~ 40 K were it not for the intervention of the SDW state with its spin-flip pair-breaking capacity. The dashed and dotted lines shown are contours in the $T$-$x_{Co}$ plane of equal relaxation rate $1/T_1T$ and of equal Knight shift $K$. The spin-lattice relaxation rate is noted to rise sharply as the SDW condition is approached. Over the rest of the plane the Knight shift data reveal a steady growth in spin pseudogapping as $T$ drops. A limiting minimum value for $K_{spin}$ is reached as $K \rightarrow K_{chem}$ = 0.225. Remember the spin component is directly proportional to the quasi-particle spin susceptibility. When presuming $\Delta_{pg}$ to be $T$ independent, it is possible to extract a pseudogap value $\approx$ 500 K or 40meV over the range where $T_c$ maximizes. [Reproduced with permission of *J. Phys. Soc. Jpn.*: copyright 2009].

**Figure 11.** (p28). Charge-, spin- and lattice-coupled RVB arrays. At the top an $S \rightarrow 0$ quadrimer array is portrayed that does not break tetragonal symmetry. In the centre are two relaxed, orthorhombic dimer twin settings. The bottom section gives an alternative orthorhombic arrangement of pair bonds taking the same basic √2x√2 cell size.

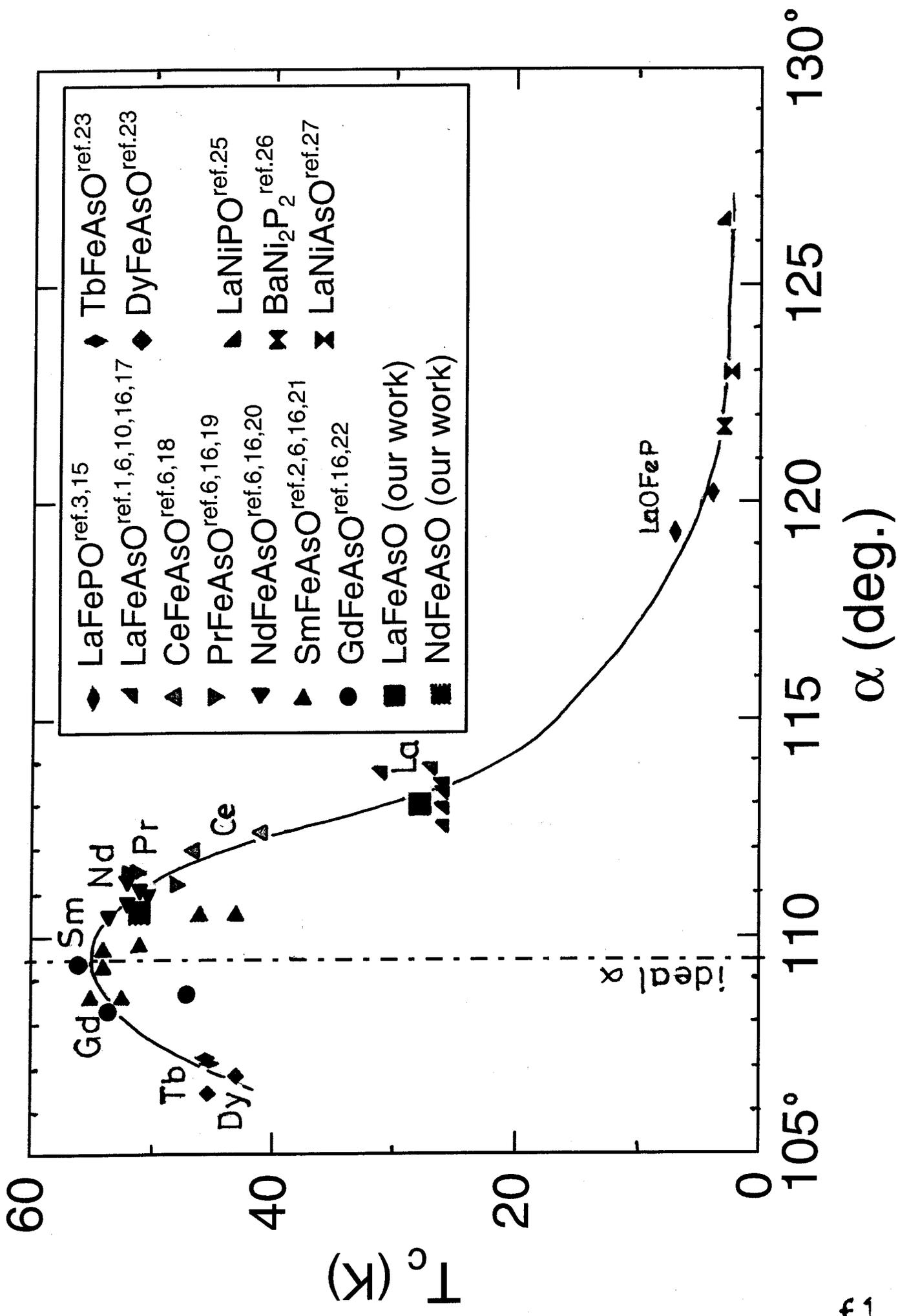

f1

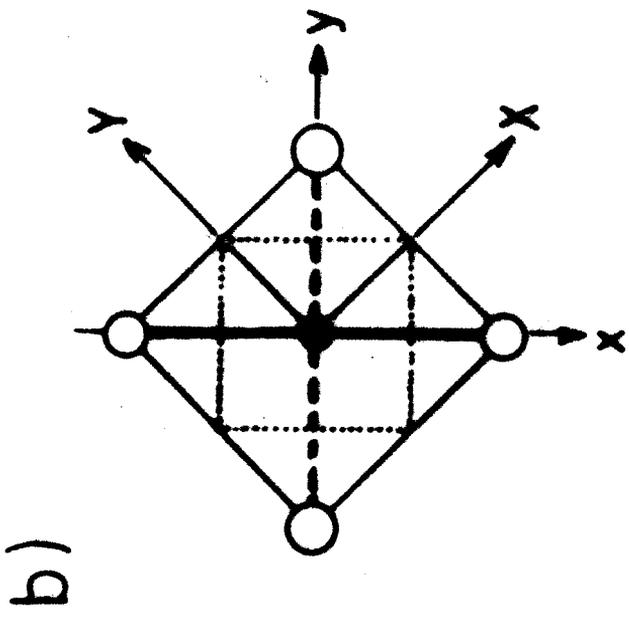

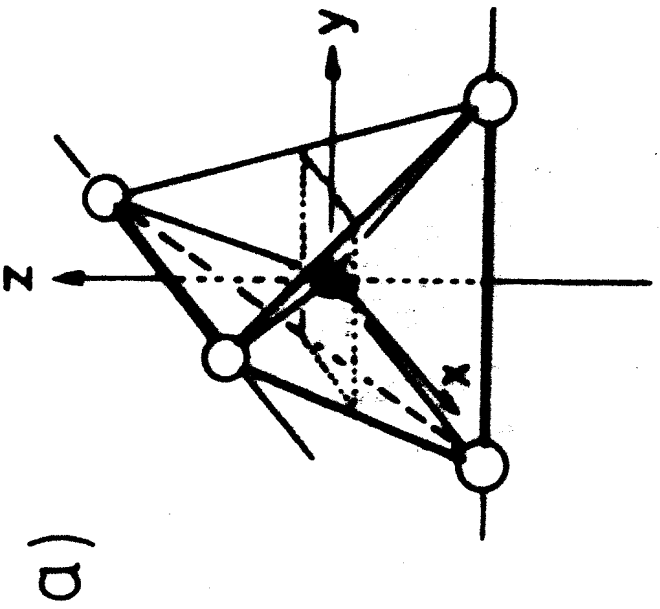

f2a,b

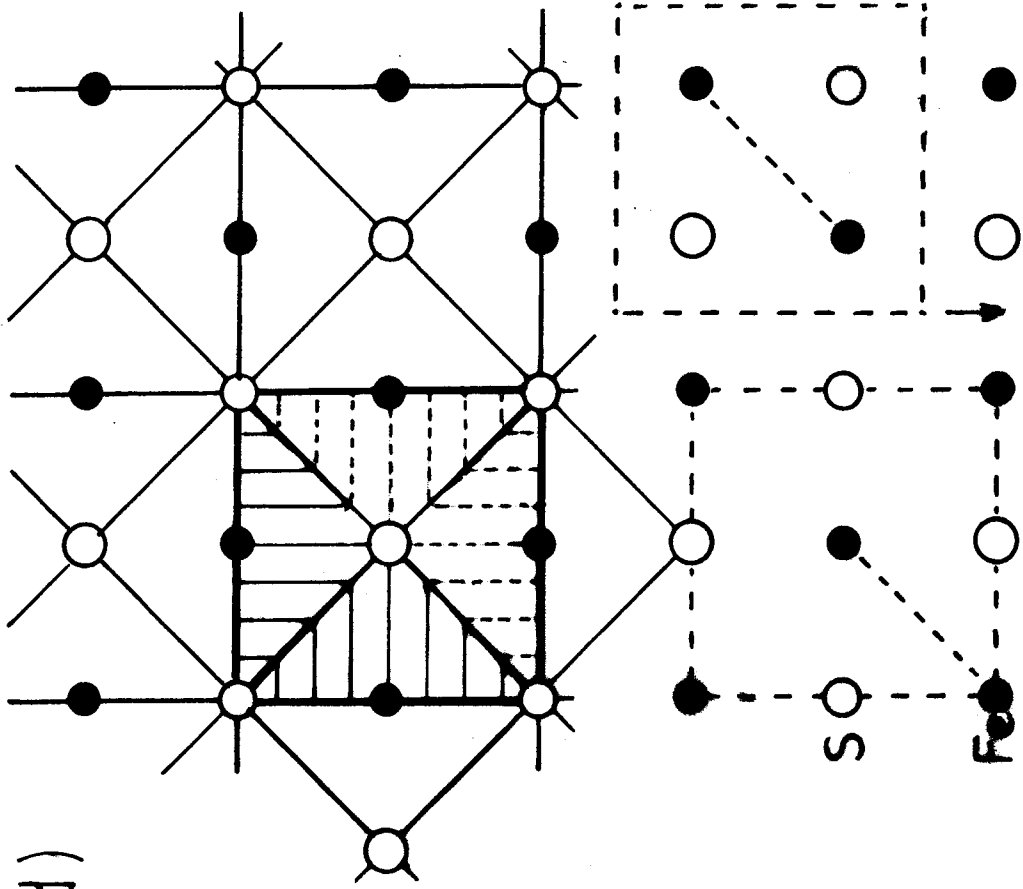

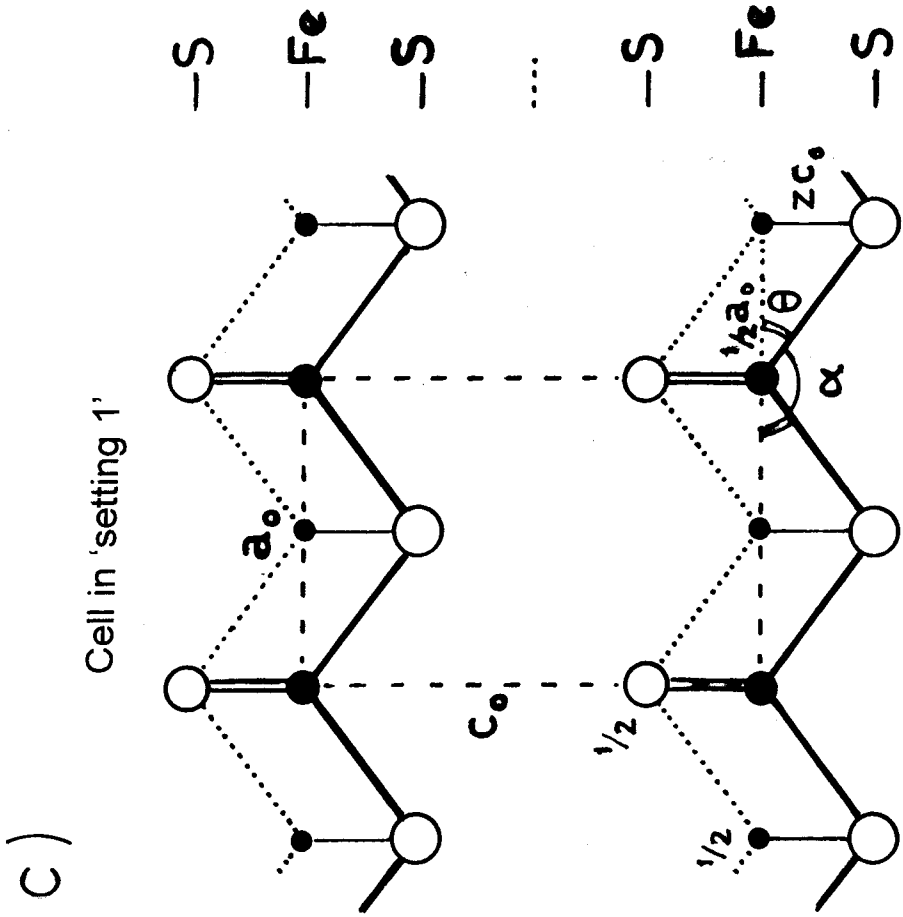

f2c,d

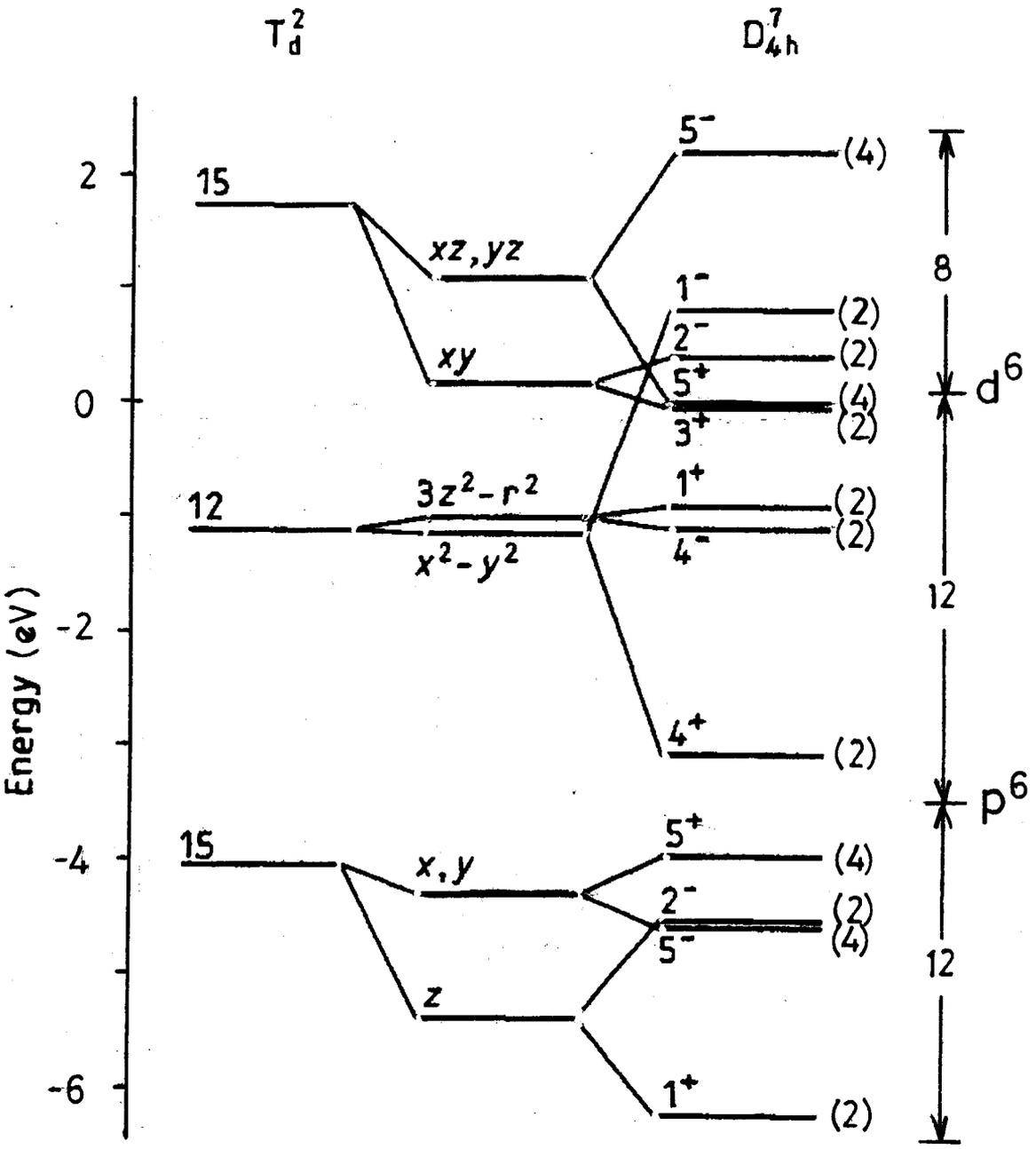

f3

Mackinowite FeS

f4

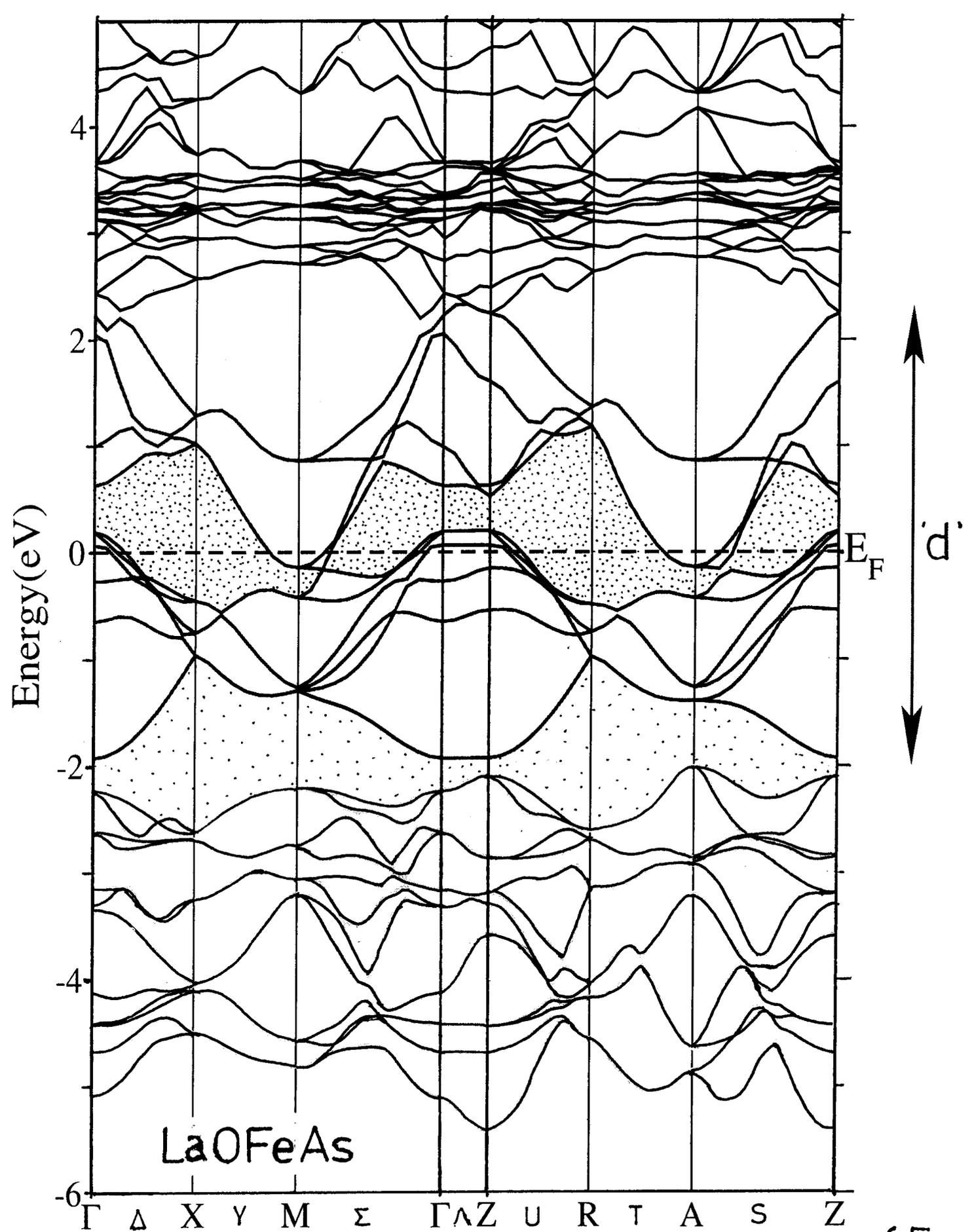

f5

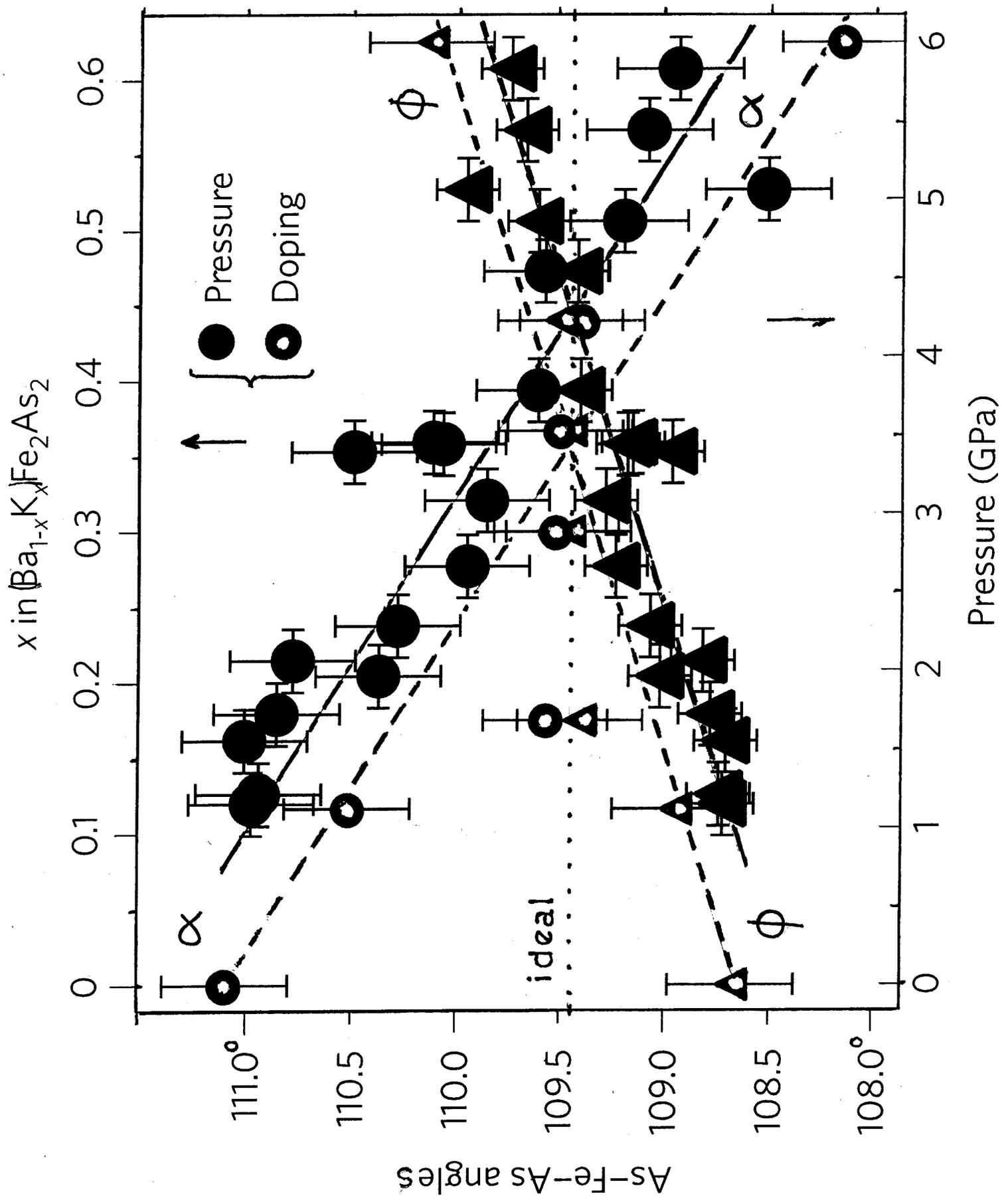

f 6

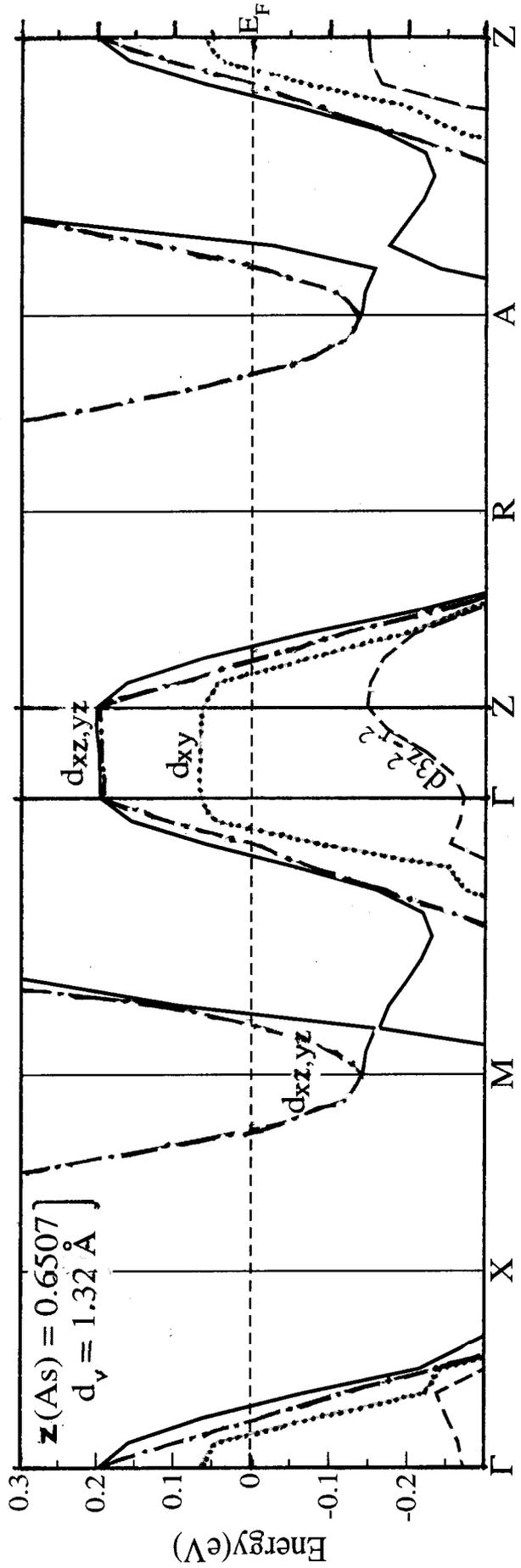
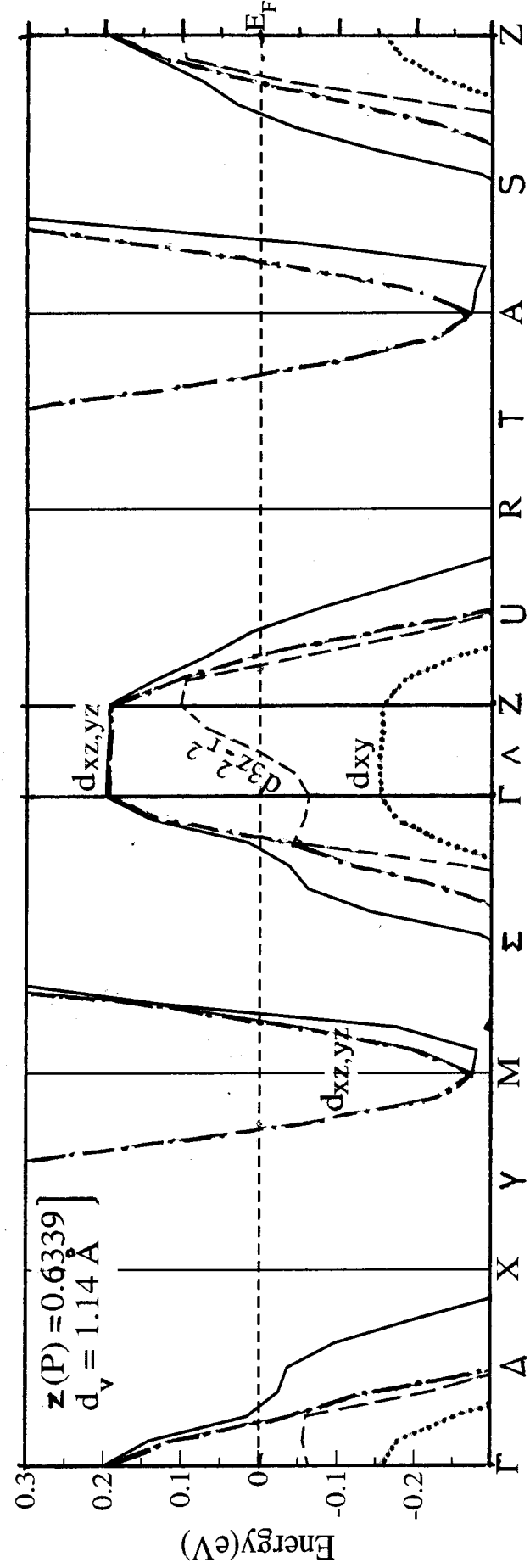

f7

(47)

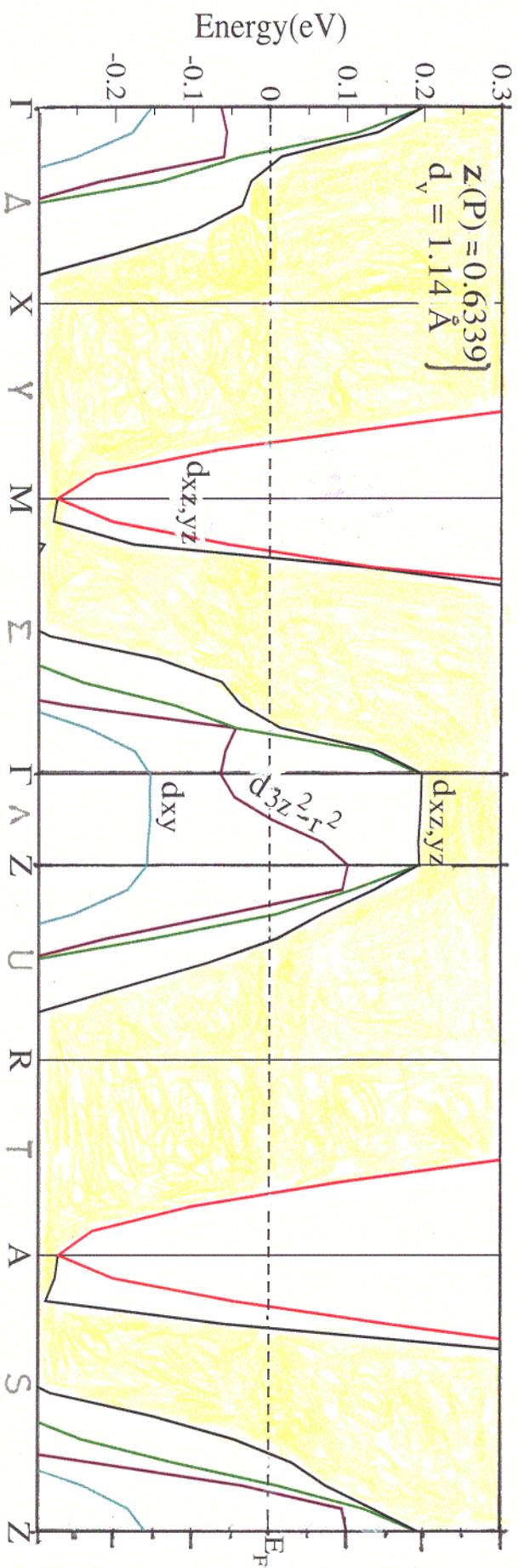

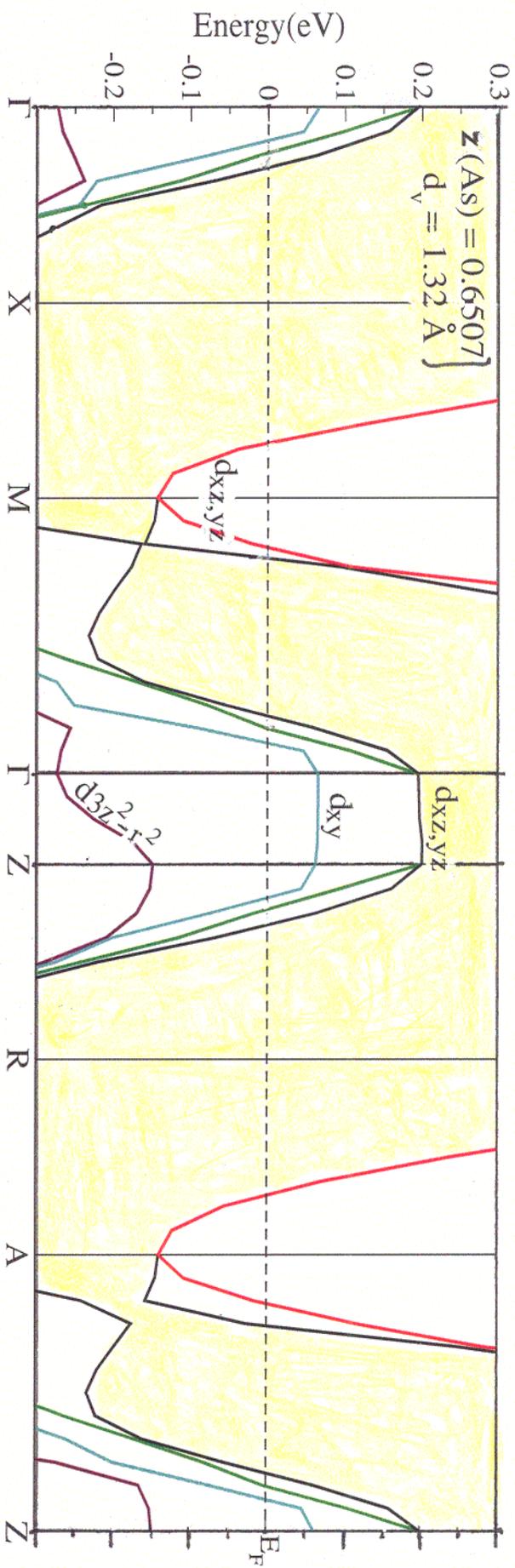

## LaOFeAs

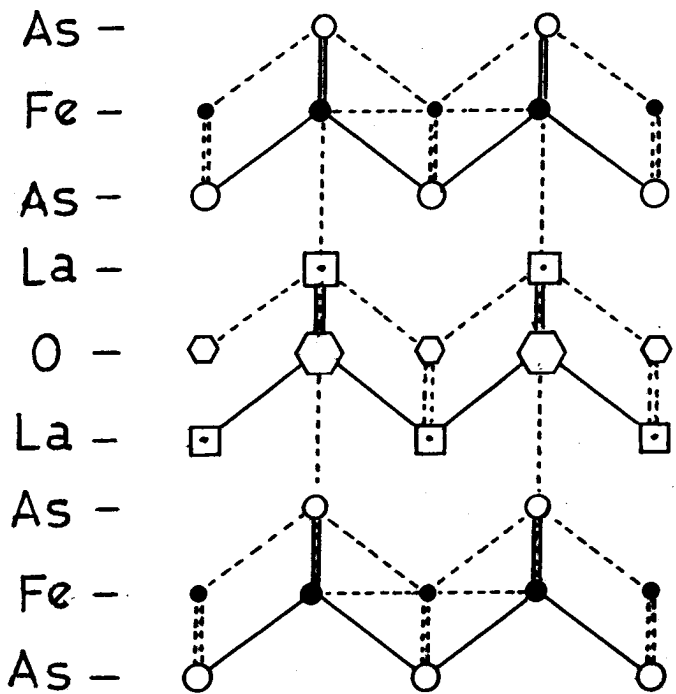

## BaFe$_2$As$_2$

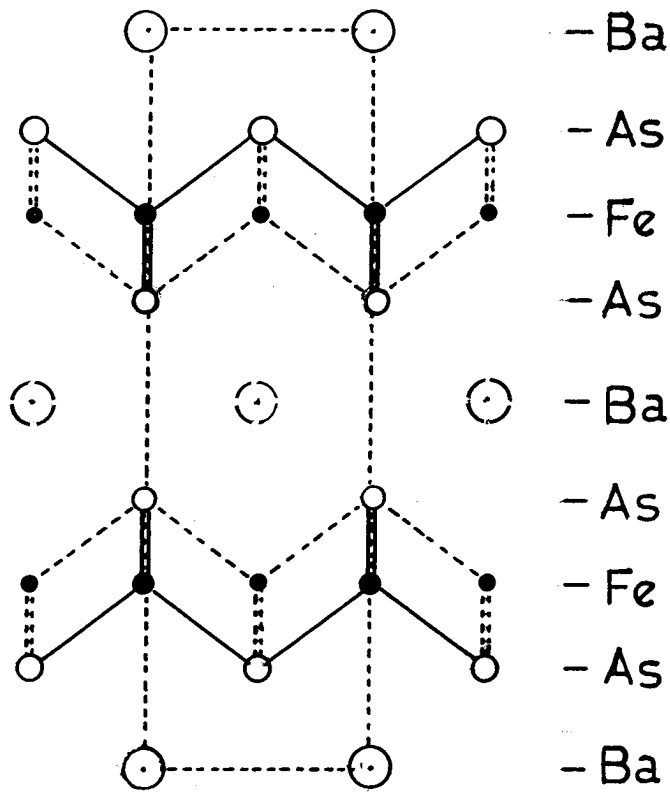

## LiFeAs

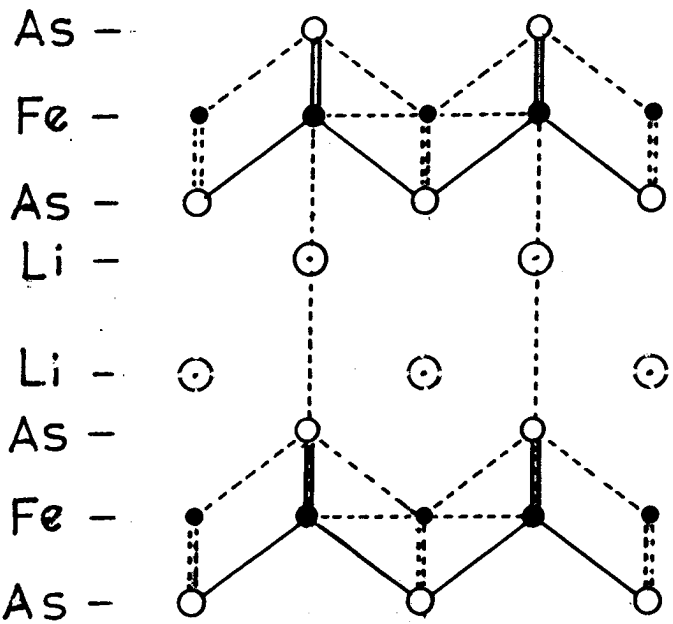

## FeSe

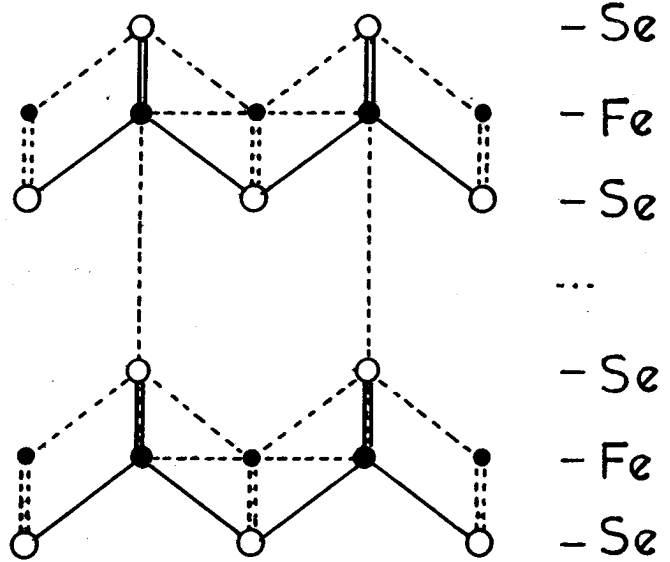

f8

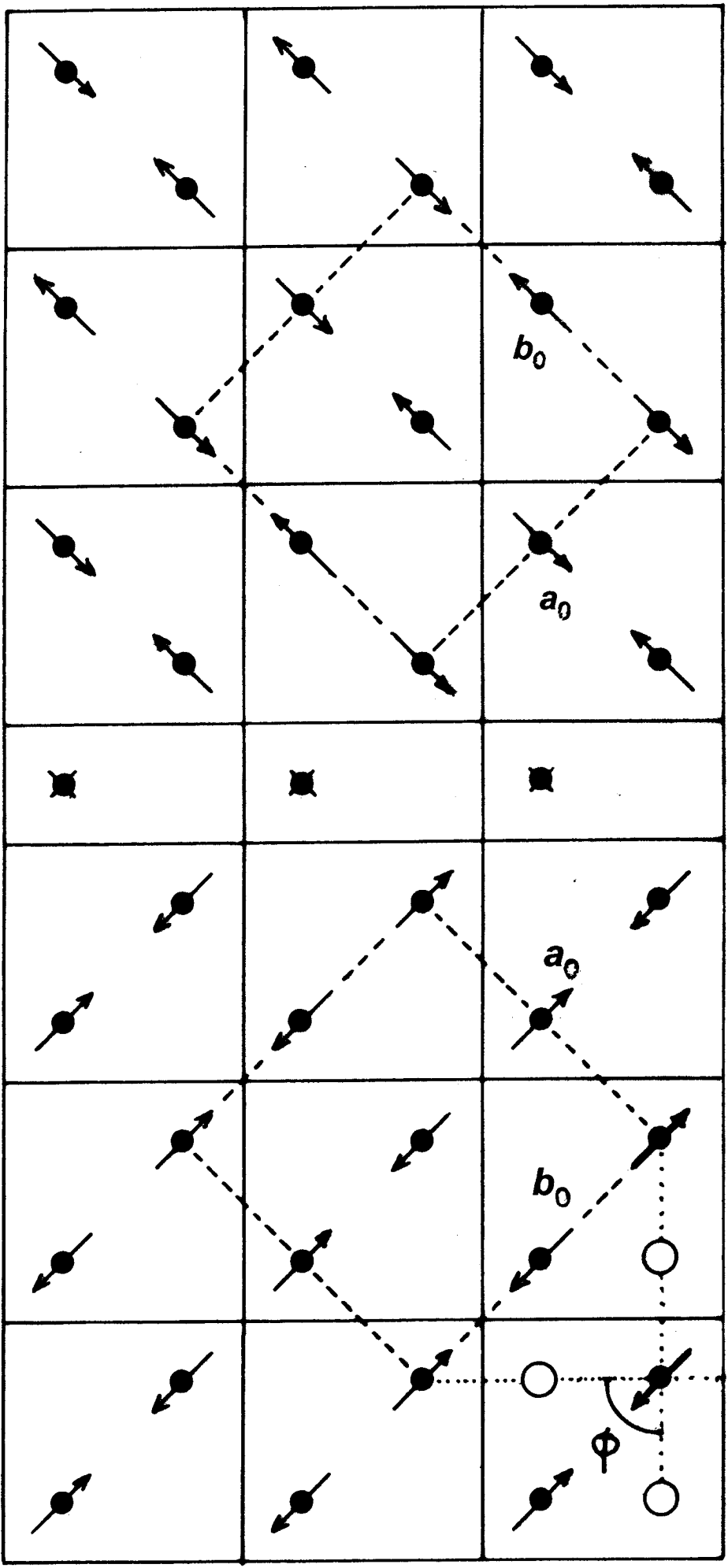

f9

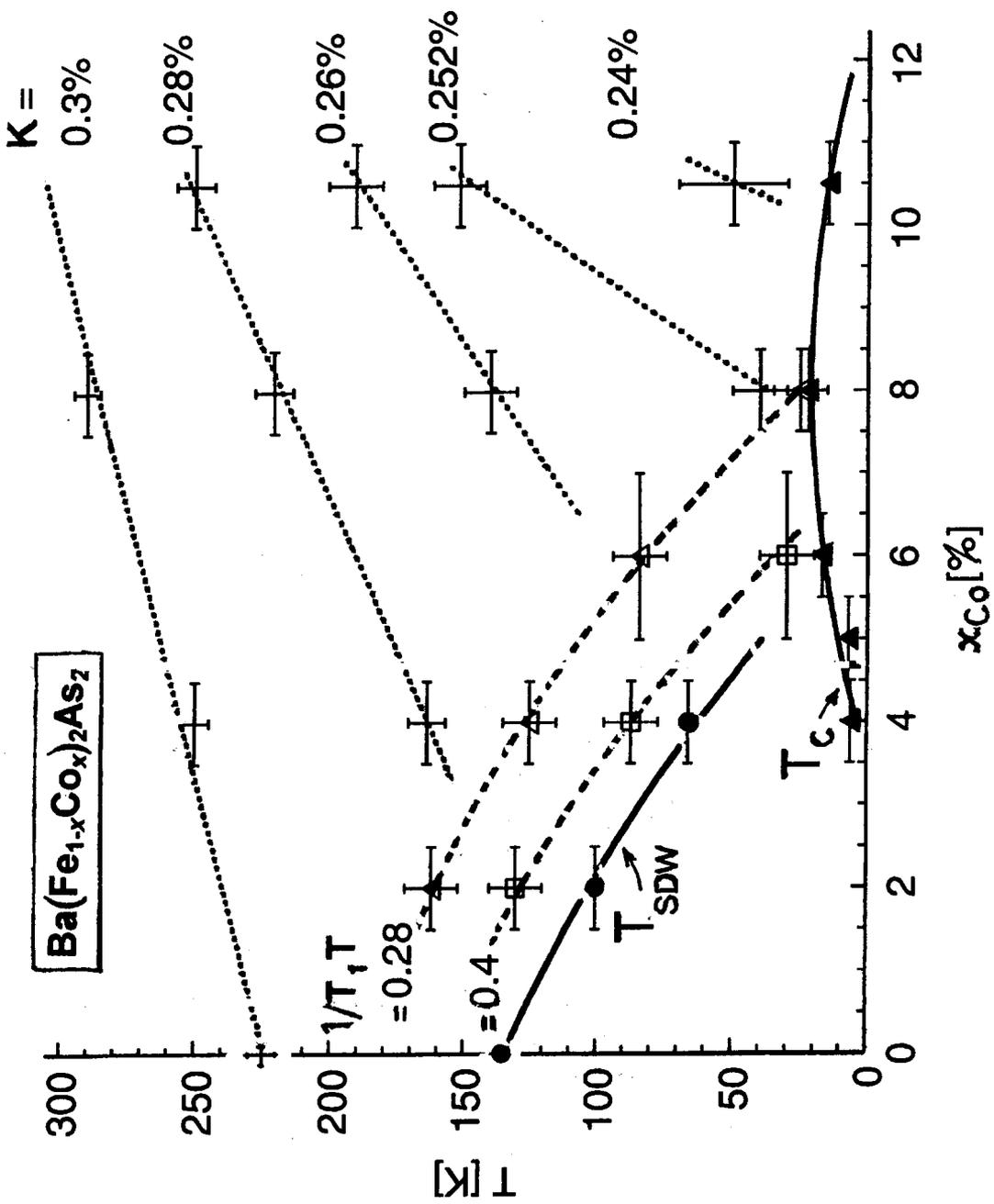

f10

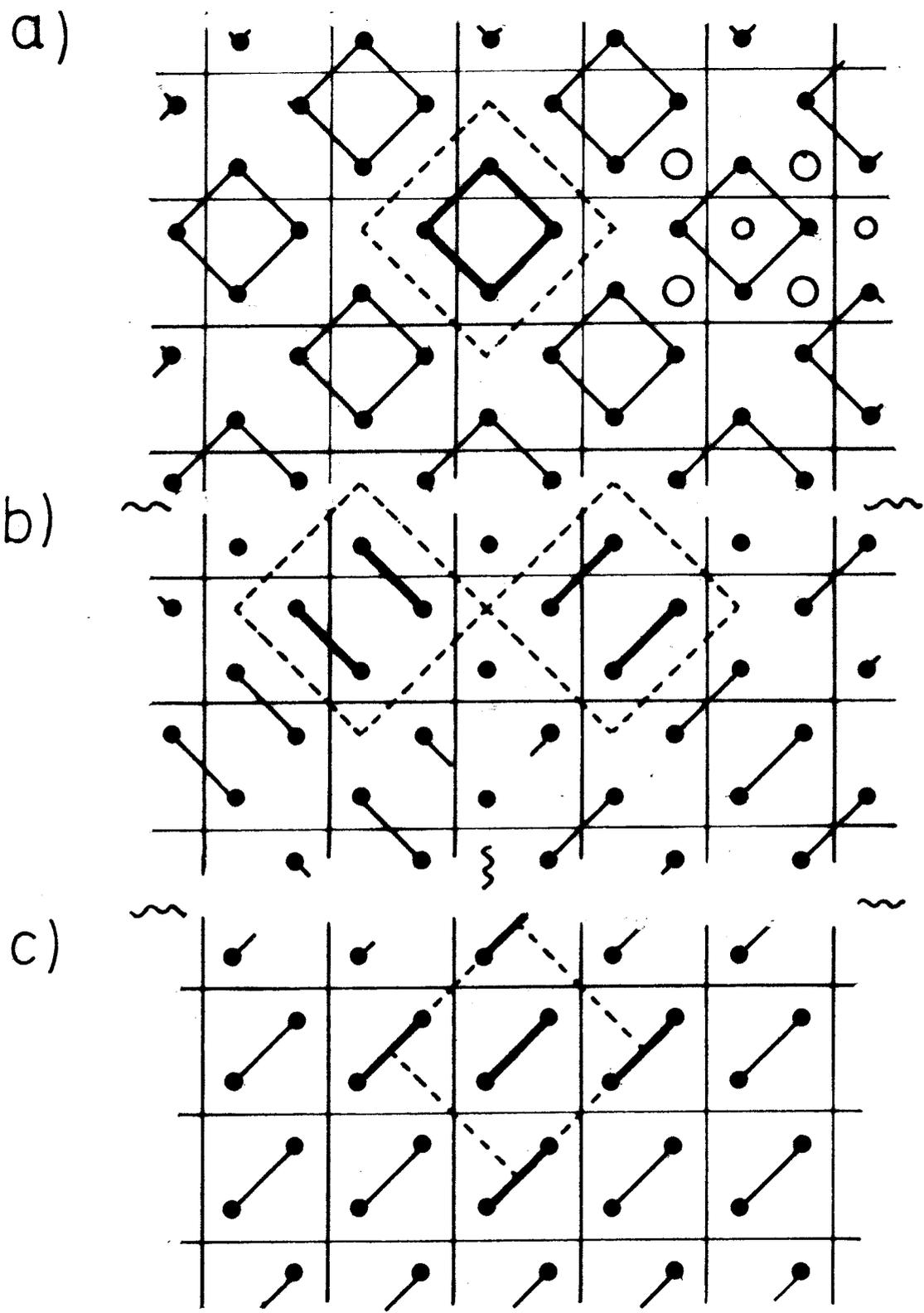

f 11